\documentclass[
aps,prd,
showpacs,
twocolumn,
notitlepage,
amssymb,amsmath,amsfonts,
nofootinbib,superscriptaddress,
floats,floatfix
]{revtex4-2}

\usepackage[colorlinks=true]{hyperref}
\hypersetup{citecolor=cyan,linkcolor=red}
\usepackage{xcolor}
\usepackage{setspace}
\newcommand{\gw}{f_{\text{gw}}}
\newcommand{\Oo}{\Omega_{\text{orb}}}

\newcommand{\nus}{\nu_{\text{s}}{}_{,\star}}
\newcommand{\spin}{\Omega_{\text{s}}}
\newcommand{\Rs}{R_{\star}}
\newcommand{\Ms}{M_{\star}}
\newcommand{\Mc}{M_{\text{comp}}}
\newcommand{\Ls}{\Lambda_{\star}}
\newcommand{\Lc}{\Lambda_{\text{comp}}}

\usepackage{graphics}
\usepackage{graphicx}
\usepackage{color}

\definecolor{maroon}{cmyk}{0,0.87,0.68,0.32}

\usepackage{journals}
	
\usepackage[]{datetime}
\begin{document}
\title{\textit{f}-mode Imprints in Gravitational Waves from Coalescing Binaries involving Aligned Spinning Neutron Stars}

\date{\today}

\author{Hao-Jui Kuan}
\email{hao-jui.kuan@uni-tuebingen.de}
\affiliation{Theoretical Astrophysics, IAAT, University of T{\"u}bingen, T{\"u}bingen, D-72076, Germany}
\affiliation{Department of Physics, National Tsing Hua University, Hsinchu 300, Taiwan}
\author{Kostas D. Kokkotas}
\affiliation{Theoretical Astrophysics, IAAT, University of T{\"u}bingen, T{\"u}bingen, D-72076, Germany}

\begin{abstract}

The excitation of $f$-mode in a neutron star member of coalescing binaries accelerates the merger course, and thereby introduces a phase shift in the gravitational waveform. 
Emphasising on the tidal phase shift by aligned, rotating stars, we provide an accurate, yet economical, method to generate $f$-mode-involved, pre-merger waveforms using realistic spin-modulated $f$-mode frequencies for some viable equations of state.
We find for slow-rotating stars that the dephasing effects of the dynamical tides  can be uniquely, EOS-independently determined by the direct observables (chirp mass ${\cal M}$, symmetric ratio $\eta$ and the mutual tidal deformability ${\tilde \Lambda}$), while this universality is gradually lost for increasing spin.
For binaries with fast rotating members ($\gtrsim800\text{ Hz}$) the phase shift due to $f$-mode will exceed the uncertainty in the waveform phase at reasonable signal-to-noise ($\rho=25$) and cutoff frequency of $\gtrsim400\text{ Hz}$. 
Assuming a high cutoff frequency of $10^3\text{ Hz}$ and fast ($\gtrsim800\text{ Hz}$) members, a significant phase shift of $\gtrsim100$ rads has been found.
For systems involving a rapidly-spinning star (potentially the secondary of GW190814), neglecting $f$-mode effect in the waveform templates can therefore lead to considerable systemic errors in the relevant analysis. In particular, the dephasing due to $f$-mode is larger than that caused by equilibrium tides by a factor of $\sim5$, which may lead to a considerably overestimated tidal deformability if dynamical tidal contribution is not accounted. The possibility of accompanying precursors flares due to $f$-mode excitation is also discussed.
\end{abstract}
\maketitle

\section{Introduction}

\subsection{The Context}

The macroscopic and microscopic properties of neutron stars (NSs) in coalescing binaries are imprinted in the emitted gravitational waveforms, thus the precise knowledge of the waveform morphology is of fundamental importance in extracting NS parameters.   
Adopting the point-particle approximation for the gravitational wave (GW) analysis, the chirp mass and the symmetric mass of binaries can be estimated from the 1st order post-Newtonian (PN) phase evolution of GWs, though with different degree of accuracy \cite{Wiseman:1992dv,Kokkotas94,Cutler:1994ys,Poisson:1995ef,Hinderer:2009ca,Damour:2012yf}. 
Beyond the point-particle baseline approximants, the internal structure of the NS members can also be probed: the quadrupole deformations induced by the tidal forces in the constituents will affect the binary evolution and thus the associated waveform. Two sorts of (gravito-electric) tidal effects are involved in the signal, viz.~the equilibrium tides due to the induced tidal deformations, and dynamical tides due to the resonant excitation of the various neutron star quasi-normal modes (QNMs). 
Equilibrium tidal effects from Newtonian \cite{Flanagan:2007ix,Hinderer:2009ca}, 1 PN \cite{Vines:2011ud,Vines:2010ca}, up to 2.5 PN level \cite{Damour:2012yf}, are encapsulated in the tidal deformability, and their traces in the signal have already been observed with the current detectors (e.g., \cite{LIGOScientific:2018cki}; see also \cite{GuerraChaves:2019foa} for a recent review). 
Dynamical tides, while being subdominant in the low frequency regime,  can affect the waveform to a similar extent as the equilibrium ones at the final stage of inspiraling, predominantly due to the $f$-mode excitation \cite{Kokkotas:1995xe,Hinderer:2016eia,Steinhoff:2016rfi,Schmidt:2019wrl}. 

For rotating NS progenitors, the spin-effects also contribute to phase shift thus introducing some degeneracy with the tidal contributions (cf.~Eq.~(3) of \cite{Dietrich:2017aum}). For instance, the spin-orbit (comes at 1.5 PN order), and  the secondary spin-spin (comes at 2 PN order) terms appear in the PN expression of GW phase \cite{Krolak:1995md,Blanchet:1995ez,Arun:2008kb}. 
In addition, rotation will induce a quadrupole deformation in stars, reshaping accordingly the form of gravitational potential.
The deformation is larger for stiffer equation of state (EOS) \cite{Dietrich:2018uni} since NSs tend to have larger radius \cite{Laarakkers:1997hb}.
The change in the gravitational potential modifies the relation between the angular velocity and the separation of stars. Binary motion is affected by this self-spin effect with leading order term coming at the same level as spin-spin effect (i.e., 2 PN) \cite{Poisson:1997ha,Levi:2014sba}. 
At leading-order, the self-spin effects in GW dephasing are however much smaller than the tidal one (see, e.g., Fig.~4 of \cite{Agathos:2015uaa}). 
Although Nagar et al. \cite{nagar19} have shown that the contribution of higher order terms of self-spin can be stronger, the enhancement is typically a factor of $\sim2$. It is thus not expected that the inclusion of these higher-order self spin effects will contribute more to the waveform dephasing than the tidal ones.

In addition, rotation indirectly modifies tidal effects by shifting and splitting QNM spectrum of NSs, which in turn affects the onset of the dynamical tides leading to waveform altering \cite{Steinhoff:2021dsn}. In particular, the downward-tuned frequencies of counter-rotating QNMs lead to stronger tidal excitations.
Although there are higher order couplings between tides and rotation, e.g., tide-spin terms, these are considerably weaker than the aforementioned effects, and even the accuracy of the state-of-the-art numerical relativity (NR) is incapable of sizing these effects \cite{Dietrich:2017aum}. 
Owing to the interplay between spin and tides, the ambiguity in the spins of inspiralling NSs would consequently obscure the determination of tidal dephasing in the gravitational waveform of binary mergers especially if one of the NSs spins rapidly (e.g., \cite{Tsokaros:2019anx,Dudi:2021wcf}). 
Concluding for the precise extraction of the source parameters, it is thus important to discriminate tidal dephasing from spin-induced phase shift.

Currently, several equation of state (EOS) candidates survive the observations of pulsars \cite{Riley:2021pdl,Raaijmakers:2021uju}, and GW170817 as well as its electromagnetic counterparts \cite{Radice:2017lry,De:2018uhw,LIGOScientific:2018cki,Fasano:2019zwm}.
Future GW detection is likely to further constrain the EOS by extracting the tidal dephasing from the waveform. Although analytic models of equilibrium tides have been derived, $f$-mode excitation has to large extent not yet been explored. Given that $f$-mode can cause a significant dephasing in some circumstances, disregarding these effects will lead to systematic errors in estimating the tidal parameters (e.g, \cite{Harry:2018hke,Pratten:2021pro}).
In a pursue of reliable analysis, it is therefore necessary to take $f$-mode effect into account when constructing gravitational waveform templates for those cases \cite{Pratten:2019sed,Pratten:2021pro,Williams:2022vct}.

The tidal effects of rotating NSs in binaries have been investigated in \cite{Steinhoff:2021dsn} by adopting an approximation for the spin modulation in frequencies of QNMs. The approximated modulation is insensitive to the EOS and the stellar mass [see Eq.~(5.7) therein], while the realistic modulation can vary by $\lesssim 15\%$ for different stars and EOS (cf.~Fig.~10 in \cite{Kuan:2021jmk}).
In the present article, we re-examine the measurability of dynamical tidal dephasing by using the realistic spin-modulation in the QNM spectrum, and a PN evolution for the inspiral part.

\subsection{This Work}

On top of a great body of existing literature, we collate in the following the original contribution of this article to address $f$-mode effect in GW:

 $\bullet$ \quad \emph{EOS-independent Hamiltonian functional} --- For slow-rotating binary NSs, the Hamiltonian governing the binary evolution, including the tidal effects, is shown to be EOS-independently reconstructable from GW observables $\mathcal{M}$, $\eta$, and $\tilde{\Lambda}$ [Eq.~\eqref{eq:uni_H}] since $f$-mode effects can be prescribed universally by $\Ls$ [Eqs.~\eqref{eq:Uni_AB}-\eqref{eq:uniB}]. Assuming we have a well-measured chirp mass \cite{Cutler:1992tc,Poisson:1995ef,Kokkotas94}, say $\mathcal{M}=1.146$ (the value for GW170817 \cite{LIGOScientific:2017vwq}), the accumulated GW phase $\Psi_{\text{tot}}$ is shown to be a universal function of the mass of the primary $\Ms$ (Fig.~\ref{fig:const_chirp}).

$\bullet$ \quad    \emph{``Observability'' of the spin effects in tidal dephasing} --- Adopting five EOS with representative spin rates, we find for symmetric binaries (masses and spins of both stars are the same) that the tidal dephasing piles up rapidly when GW has a frequency $\gw>400\text{ Hz}$ (top panel of Fig.~\ref{fig:spin}), implying that the information of dynamical tides concentrates in this late part of the waveform. 
Furthermore, the tidal dephasing accumulated from $\gw=20$ to 1000 Hz is larger for higher stellar spins and/or stiffer EOS  (bottom panel of Fig.~\ref{fig:spin}). For the stiffest EOS considered, MPA1, we find a few tens of rads of dephasing if two stars rotate moderately, while a few hundreds of rads can be accumulated when stars rotate rapidly. 
Although we consider symmetric binaries in Fig.~\ref{fig:spin}, the conclusions are expected to be general since (i) mode frequencies are reduced further by faster rotation, enhancing tidal dephasing, and (ii) NSs with stiffer EOS tend to have larger radii thus more notable tidal deformations. Despite of its dependence on EOS, the tidal dephasing can be expressed as a universal relation with respect to a dimensionless spin [Fig.~\ref{fig:uni_psi_spin}; Eq.~\eqref{eq:uni_psi_spin}].

  $\bullet$ \quad  \emph{Fast-spinning NS} --- In general, the signal-to-noise (SNR) needed to measure the tidal dephasing $\Delta\Psi^T$, which is defined as $\rho_{\text{thr}}$, depends on the cutoff frequency of the data stream $f_{\text{max}}$, and the spin of the primary $\spin{}_{,\star}=2\pi\nus$. 
Taking a specific binary as an example, we see that the error in GW phase decreases, while the tidal dephasing increases for larger $f_{\text{max}}$ (top panel of Fig.~\ref{fig:compare}). To emphasise how these two parameters affect the measurability of tidal dephasing, we plot $\rho_{\text{thr}}$ as a function of $\nus$ for four representative cutoffs. We see that $\rho_{\text{thr}}$ decreases logarithmically as $f_{\text{max}}$ is linearly increased; for a reasonable SNR $\rho=25$, the tidal dephasing may be detected only for $\nus>600$ Hz if the cutoff is 400 Hz, while a spin $>400$ Hz may already be detectable if the cutoff is 100 Hz higher. 

Using our approach, the tidal dephasing can be observed for the SNR of the event GW190814 $\rho\lesssim20$ if (i) the secondary of GW190814 is a fast-rotating NS with $\nus>800\text{ Hz}$, (ii) the star almost aligns with the orbit, and (iii) the waveform is observed up to merger, which occurs at $\sim360$ Hz (Sec.~\ref{case0814}).
In other words, the nature of the secondary of GW190814, which may have been either the heaviest NS or the lightest black hole that has ever be seen, may be determined by hunting the tidal dephasing.
\emph{The sizable $f$-mode effects quantified in the present article indicate that templates including $f$-mode effects are imperative for inferring EOS from more accurate GW observation in the near future.}
That said, if a rapidly-rotating NS is involved in a coalescing binary, as it is claimed for the secondary of GW190814 \cite{Zhang:2020zsc,Most:2020bba,Biswas:2020xna}, the $f$-mode effects enhanced by the fast rotation will be unambiguously measurable in the signal, which may even be more significant than the imprint of equilibrium tides.
Although neglecting $f$-mode excitations will not affect noticeably the estimates of the chirp mass and symmetric mass ratio since the overall tidal effects in waveform are minor to the influences of these two parameters, the inference of tidal parameters, and in turn the EOS, will be significantly biased.

In addition, the $f$-mode excitation in fast-rotating stars may generate enough strain to yield the crust. If certain conditions are met, some flares may be launched before merger. These precursors are however not associated with the resonance shattering scenario \cite{Tsang:2011ad,Suvorov:2020tmk,Kuan:2021sin}. Instead, non-resonantly-excited $f$-modes may yield the crust owing to their strong coupling to the exterior tidal field. We find for the specific case of GW190814, such precursor may be emitted at $\sim0.6$ s prior to merger. The non-observation does not necessarily exclude the NS-nature of the lighter member since several factors could affect the formation of precursor flares.

The article is organised as following: We summarise the present status of the analytical waveform derived from the effective-one-body approach in Sec.~\ref{pnphase}. Tidal dephasing is numerically studied in Sec.~\ref{simu}, and its reliability is tested against the waveform models discussed in Sec.~\ref{pnphase}. The dependence on EOS, and the tidal effects of spinning NS are also detailed there. 
The dephasing due to $f$-mode excitation and its influence in the waveform analysis are investigated for GW190814-like events in Sec.~\ref{case0814}, where we also assess on the possibility of electromagnetic counterparts.
A discussion of the results is provided in Sec.~\ref{discussion}.

Unless stated otherwise, all quantities are given in the unit of $c=1=G$.

\section{Tidal dephasing: Analytic Models}\label{pnphase}

Under the stationary phase approximation (SPA) and by ignoring the PN modifications in the GW amplitude $\mathcal{A}$\footnote{In general, the amplitude $\mathcal{A}(\gw)$ depends on the internal structure (or finite size effects) of the binary members,  the omission of these higher PN corrections does not affect the accuracy of SPA. In practice, SPA will be quite accurate up to the merger \cite{Damour:2012yf,Droz:1999qx}, and thus this approximation will not affect significantly our results.}, the frequency-domain gravitational waveform can be expressed by: \cite{Cutler:1994ys,Poisson:1995ef}
\begin{align}\label{eq:SPA}
	h(\gw)=\mathcal{A}\gw^{-7/6} e^{i\Psi(\gw)}\, .
\end{align}
This form is generic for all kinds of compact binaries for $\gw=\Oo/\pi$ the GW frequency, and $\Oo$ the orbital frequency. 
The phase $\Psi(\gw)$ is related to the time-domain phase $\phi(t)$ via
\begin{align}\label{eq:psi_lagendre}
	\Psi(\gw) = 2\pi\gw t_o -\phi(t_o) -\frac{\pi}{4},
\end{align}
and obeys the equation \cite{Baiotti:2011am,Damour:2012yf,Schmidt:2019wrl}
\begin{align}
	\frac{d^2\Psi}{d\Oo^2}= \frac{2Q_{\omega}}{\Oo^2},
\end{align}
where $t_o$ is a reference time with $\phi_o=\phi(t_o)$ being the corresponding phase, and $Q_{\omega}$ is a dimensionless measure of the phase acceleration defined as
\begin{align}\label{eq:phiacc}
	Q_{\omega}=\Oo^2\bigg(\frac{d\Oo}{dt}\bigg)^{-1}.
\end{align}

The construction of precise waveforms demands accurate evolution of coalescing binaries. The PN equations of motion describe most part of the observed inspiralling evolution with adequate accuracy \cite{Blanchet:2013haa}. Still the PN approximation gradually fails as the binary approaches the plunge, merger, and finally the ringdown phases. 
This lack of applicability for high orbital frequencies and the post-merger dynamics motivated the so-called effective-one-body (EOB) formalism, which re-sums the PN expansions to account properly for the higher-order effects \cite{Buonanno:1998gg,Damour:2009kr}. 
In addition, the EOB analytic dephasing can be ``calibrated'' with the late-time NR results \cite{Damour:2012ky} even when tidal effects are taken into account \cite{Kawaguchi:2018gvj,Dietrich:2019kaq}. 
However, the latter hybrid EOB and NR model is significantly more time-consuming than the PN formalism.
In the present article, we aim to offer an economical waveform variant capable to probe the internal physics of NSs from pre-merger waveforms ($\lesssim10^3$ Hz). To this end, the PN framework proves sufficiently accurate \cite{Boyle:2007ft,Hotokezaka:2015xka,Bernuzzi:2014owa} (see also below).

Tidal interactions among the binary members  
(i) accelerate the shrinking rate due to  orbital energy transfer to QNMs \cite{Lai:1993di,Kokkotas:1995xe}, or, in another perspective, effectively amplify the strength of the gravitational potential in the EOB framework \cite{Damour:2009wj,Bini:2012gu,Bernuzzi:2014owa}, 
(ii) enhance the energy flux carried by GW due to tidal deformations (see, e.g., Eq.~(3.6) of \cite{Vines:2011ud}), and 
(iii) increase  the  angular momentum loss. 
As a result, certain tidally-driven modifications in $\mathcal{A}$ and $\Psi$ will be encoded, while the change in $\mathcal{A}$ in minor.

The effect of equilibrium tides in the phase shift is mainly governed by the quadrupolar tidal deformability of the NSs in the binary \cite{Hinderer:2009ca,Flanagan:2007ix},
\begin{align}
	\Lambda = \frac{2k_2}{3C^5}
\end{align}
 where $k_2$ is the (dimensionless, quadrupolar) tidal Love number, and $C=M/R$ is the stellar compactness. 
 The contribution of the higher-order Love numbers in the phase-shift is significantly smaller \cite{Dietrich:2017aum}.
 On the other hand, the dynamical tides are due to the excitation of oscillations in the individual NSs, predominantly by the quadrupolar ($l=2$) component of the tidal potential built by the companion (see, e.g., the discussion in the Appendix A.2 of \cite{Damour:2012yf}). 
Among the various low or high frequency modes ($p$-, $g$-, $i$-, $w$-, etc.), the tidal response of the $l=2$ $f$-mode is much more significant \cite{Kokkotas:1995xe,Damour:2012yf,Lackey:2016krb,Dietrich:2017aum}. 
To study the leading order tidal phenomena, we restrict ourselves to the physics ($f$-mode, tidal potential and deformability) at the quadrupolar level. Our methodology is obviously applicable to higher-order ($l>2$) $f$-modes as well as other types of modes.

Stellar spins also influence the binary evolution via spin-orbit, self-spin couplings \cite{Kidder:1992fr,1994PhRvD..49.6274A,Krolak:1995md,Blanchet:1995ez}, and some higher order terms such as spin-tidal coupling. 
Incorporating the aforementioned physics, and by denoting a certain parameter $X$ of the primary (companion) as $X_{\star}$ ($X_{\text{comp}}$), the GW phase can be expressed as
\begin{align}
	\Psi=\Psi(\gw;\mathcal{T},\mathcal{S},\mathcal{Z}),
\end{align}
where $\mathcal{T}=(\Ls,\Lc)$, $\mathcal{S}=(\nus,\nu_{\text{s,comp}})$, and $\mathcal{Z}=(\Ms,\Mc,R_{\star},R_{\text{comp}})$. 
In the present article, however, we focus on the influence of spins in the dynamical tides, and we will ignore other spin-related dephasing such as spin-orbit, spin-spin, self-spin, etc.

\subsection{Analytic Tidal Dephasing}\label{analtidphi}

Equilibrium tides are usually addressed by extending the effective gravitational potential in EOB to include an enhancement due to higher-order PN contributions \cite{Damour:2009wj,Bernuzzi:2014owa}, while dynamical tides can be investigated either by introducing associated kinetic terms to the Hamiltonian \cite{Alexander:1987zz,Steinhoff:2016rfi} or by generalising the Love number $k_2$ to a running parameter (effective tidal responses) \cite{Maselli:2012zq,Hinderer:2016eia,Steinhoff:2016rfi}.
We adopt the PN evolution of binaries together with a kinetic term for dynamical tides to investigate the accumulated  tidally-induced dephasing during the pre-merger stages ($\gw\lesssim10^3$ Hz). This approach is numerically cheaper  while agrees very well with the more sophisticated EOB method (cf.~Fig.~1 in \cite{Schmidt:2019wrl}; see also Fig.~\ref{fig:comparison_psi}). 
To demonstrate the faithfulness of our code, we will compare our results with several analytical waveforms in this section. 

We begin with a brief summary of the waveform models that we are going to compare to.
The 1 PN order phase shift due to the equilibrium tides in the primary, based on the \text{TaylorF2} model, reads as \cite{Vines:2011ud} 
\begin{align}\label{eq:1pnshift}
	\Delta\Psi^{\text{eq}}_{\star}
	&=-\frac{3\Ls}{128}(\pi\mathcal{M}\gw)^{-5/3}x^{5}[a_0+a_1x] \nonumber\\
	&\approx \Psi(\gw;\mathcal{T},\mathcal{S},\mathcal{Z})-\Psi(\gw;\mathcal{O},\mathcal{S},\mathcal{Z}),
\end{align}
where $x=[\pi (\Ms+\Mc) \gw]^{2/3}$, $\mathcal{O}=(0,0)$ denotes a null pair. The chirp mass and the symmetric mass ratio are defined, respectively, by
\begin{align}
	\mathcal{M} = \frac{(\Ms \Mc)^{3/5}}{(\Ms+\Mc)^{1/5}}, \,\, \text{and}\,\,
	\eta = \frac{\Ms\Mc}{(\Ms+\Mc)^2}.
\end{align}
The coefficient
\begin{subequations}
\begin{align}
	a_0 =& 12[1+7\eta-31\eta^2-\sqrt{1-4\eta}(1+9\eta-11\eta^2)],\\
	\intertext{is the Newtonian contribution, and}
	a_1 =& \frac{585}{28}\bigg[ 1+ \frac{3775}{234}\eta-\frac{389}{6}\eta^2+\frac{1376}{117}\eta^3 \nonumber\\
	&-\sqrt{1-4\eta}\bigg(1+\frac{4243}{234}\eta-\frac{6217}{234}\eta^2-\frac{10}{9}\eta^3\bigg)\bigg],
\end{align}
\end{subequations}
is the 1~PN one.
Although Eq.~\eqref{eq:1pnshift} encodes solely the tide in the primary, the effects from the companion can be linearly added to the gross influence, which can be simplified as \cite{Favata:2013rwa,Wade:2014vqa}
\begin{align}\label{eq:1PNTaylorF2}
	\Delta\Psi^{\text{eq}}_{\star}&+\Delta\Psi^{\text{eq}}_{\text{comp}}=
	-\frac{3\tilde{\Lambda}}{128}(\pi\mathcal{M}\gw)^{-5/3}x^{5} \nonumber\\
	& \times\bigg[ \frac{39}{2} +\bigg(
	\frac{3115}{64}-\frac{6595}{364}\sqrt{1-4\eta}\frac{\delta\tilde{\Lambda}}{\tilde{\Lambda}} \bigg) x
	\bigg],
\end{align}
where
\begin{align}
	\tilde{\Lambda} 
	=&\frac{16}{13(\Ms+\Mc)^5}\bigg[(\Ms+12\Mc)\Ms^4\Ls\nonumber\\
	&+(\Mc+12\Ms)\Mc^4\Lc\bigg]
\end{align}
is the mutual tidal deformability, and
\begin{align}
	\delta\tilde{\Lambda}=&\frac{1}{2}\bigg[\sqrt{1-4\eta}\bigg(1-\frac{13272}{1319}\eta+\frac{8944}{1319}\eta^2\bigg)(\Ls+\Lc)\nonumber\\
	&+\bigg(1-\frac{15910}{1319}\eta+\frac{32850}{1319}\eta^2+\frac{3380}{1319}\eta^3\bigg)(\Ls-\Lc)
	\bigg].
\end{align}
We note that $\delta\tilde{\Lambda}$ is typically much smaller than $\tilde{\Lambda}$, and will vanish for symmetric binaries.

The above 1 PN form can be extrapolated to 2.5 PN with the aid of EOB treatment as shown in \cite{Damour:2012yf}; for symmetric binaries, the equilibrium tidal effect of the primary leads to the dephasing
\begin{align}\label{eq:DTaylorF2}
	\Delta\Psi_{\star}^{\text{eq}}=&-\frac{117\Ls}{128}x^{5/2}\bigg[1+\frac{3115}{1248}x-\pi x^{3/2}\nonumber\\
	&+\frac{28024205}{3302208}x^{2}-\frac{4283}{1092}\pi x^{5/2}\bigg],
\end{align}
and can be doubled to include the influence of the companion. 
The 2.5 PN correction to the phase acceleration \eqref{eq:phiacc},
\begin{align}\label{eq:PN_Q}
	\tilde{Q}_{\omega}^{T} =& -\frac{65}{32} x^{5/3}\bigg[1+\frac{4361}{624}x^{2/3}-4\pi x\nonumber\\
	&+\frac{5593193}{122304}x^{4/3}-\frac{4283}{156}\pi x^{5/3}\bigg],
\end{align}
is also given in \cite{Damour:2012yf}. We will compare our results to analytic forms of tidal dephasing and of phase acceleration. 
We note that there is a mutation of 2.5 PN tidal phase approximant derived by Henry et al. \cite{Henry:2020ski}, which is slightly different from the one in \cite{Damour:2012ky}. However, they match up to 1 PN tidal phasing.

This 2.5 PN TaylorF2 model was later phenomenologically calibrated by numerical relativity simulations in \cite{Kawaguchi:2018gvj} to capture dynamical tidal effects by replacing $\Lambda$ in Eq.~\eqref{eq:DTaylorF2} with $\Lambda(1+12.55\Lambda^{2/3}x^{4.240})$. Ignoring the contribution of $\delta\tilde{\Lambda}$, the authors further show that one can directly generalise the above phase expression, which is for a single star in symmetric binaries, to that for both stars in asymmetric systems by substituting the denominator of the pre-factor with $256\eta$ and changing $\Ls$ to $\tilde{\Lambda}$, i.e.,
\begin{align}\label{eq:kawaguchi}
	\Delta\Psi_{\star}^{\text{eq}}&+\Delta\Psi_{\text{comp}}^{\text{eq}}=-\frac{117}{256\eta}x^{5/2}\tilde{\Lambda}(1+12.55\tilde{\Lambda}^{2/3}x^{4.240})\nonumber\\
	&\times\bigg[1+\frac{3115}{1248}x-\pi x^{3/2}+\frac{28024205}{3302208}x^{2}-\frac{4283}{1092}\pi x^{5/2}\bigg].
\end{align}
As will be shown in Fig.~\ref{fig:comparison_psi}, this NR-reshaped, TaylorF2 model matches well to our numerical results, thus we will use the above NR-reshaped form for statistical estimation in Sec.~\ref{stat.err}.
In addition to TaylorF2, the NR-calibrated form for TaylorT2 has also been derived in \cite{Dietrich:2017aum}, which gives rise to the dephasing
\begin{align}\label{eq:dietrich}
	\Delta\Psi_{\star}^{\text{eq}}&+\Delta\Psi_{\text{comp}}^{\text{eq}}=-\frac{117\tilde{\Lambda}}{64}x^{5/2}\nonumber\\
&\times\frac{1-17.428x+31.867x^{3/2}-26.414x^2+62.362x^{5/2}}{1-19.924x+36.089x^{3/2}}.
\end{align}

Although several models for equilibrium tides have been developed, dynamical tidal dephasing due to the quadrupolar $f$-mode in the primary is derived recently by Schmidt \& Hinderer \cite{Schmidt:2019wrl}, and is given by
\begin{align}\label{eq:fmode}
	\Delta\Psi^{\text{dyn}}_{\star} =& -\frac{10\sqrt{3}\pi-27-30\log2}{96\eta}(\pi \Ms \gw)^{11/3}\nonumber\\
	&\times\frac{\Ls\Ms^{4}\omega_f^{-2}}{(\Ms+\Mc)^6}\bigg(155-147\frac{\Ms}{\Ms+\Mc}\bigg),
\end{align}
where $\omega_f$ is the frequency of the $f$-mode. This analytical phase shift agrees with the tidal EOB model for $\gw\lesssim 10^3$ Hz \cite{Schmidt:2019wrl}.

\section{Tidal Dephasing: Numerical results}\label{simu}

The Hamiltonian describing the binary evolution is \cite{Alexander:1987zz,Kokkotas:1995xe,Kuan:2021jmk}
\begin{align}\label{eq:ham}
	H(t)=\big(H_{\text{orb}} + H_{\text{reac}} + H_{\text{osc}} + H_{\text{tid}} \big) (t),
\end{align}
where $H_{\text{tid}}$ and $H_{\text{osc}}$ are the Hamiltonians for the equilibrium and dynamical tidal effects, respectively. We consider the conservative motion, $H_{\text{orb}}$, up to 3 PN level, and include the gravitational back-reaction, $H_{\text{reac}}$, at 2.5 PN order \cite{Kuan:2021jmk}. 
The explicit forms for the point-particle part of $H(t)$, viz.~$H_{\text{orb}}$ and $H_{\text{reac}}$, are rather long and are not the subject of the present article. We thus omit them here but we refer the interested reader to \cite{Schaefer:1985vxb,Damour:1999cr}. Nonetheless, we will provide the form of the tidal parts $H_{\text{osc}}+H_{\text{tid}}$ in Sec.~\ref{UR}, for which the coupling strength of modes to the external tidal field is a crucial parameter. 

\subsection{Numerical setup and validation}

The binary evolution is obtained by solving the Hamiltonian equations associated to the Hamiltonian \eqref{eq:ham} (see, e.g., Sec.~3 of \cite{Kuan:2021jmk} for explicit expressions), where the initial separation is set for a binary in a quasi-circular orbit with initial orbital frequency at 10 Hz. Using the 4th order Runge-Kutta method with a time step of $5\times10^{-4}P_\text{orb}$ for orbital period $P_\text{orb}$, we evolve the binary up to $\gw=1000$~Hz. The resolution has been shown to be adequate for achieving numerically converging results.

\begin{figure}
	\centering
	\includegraphics[width=\columnwidth]{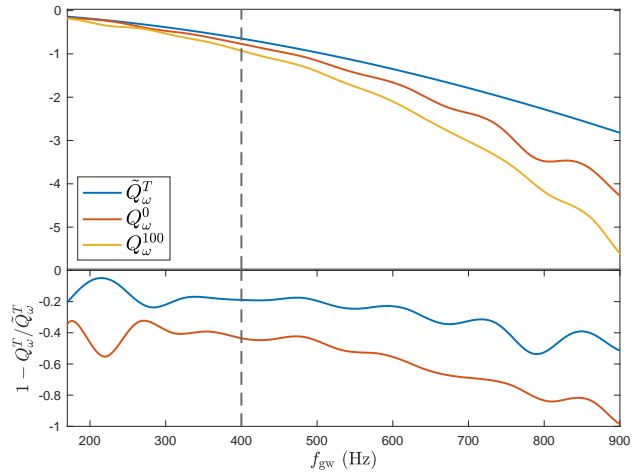}
	\caption{Numerically estimated phase acceleration $Q_{\omega}^T=\{Q_{\omega}^0,Q_{\omega}^{100}\}$ [Eq.~(4)] and the analytic 2.5 PN form $\tilde{Q}_{\omega}^{T}$ [Eq.~(15)] (top panel). The relative deviation between $Q_{\omega}^{T}$ and $\tilde{Q}_{\omega}^{T}$ is shown in the bottom panel.
	}
	\label{fig:comparison_Q}
\end{figure}

\begin{figure}
	\centering
	\includegraphics[width=\columnwidth]{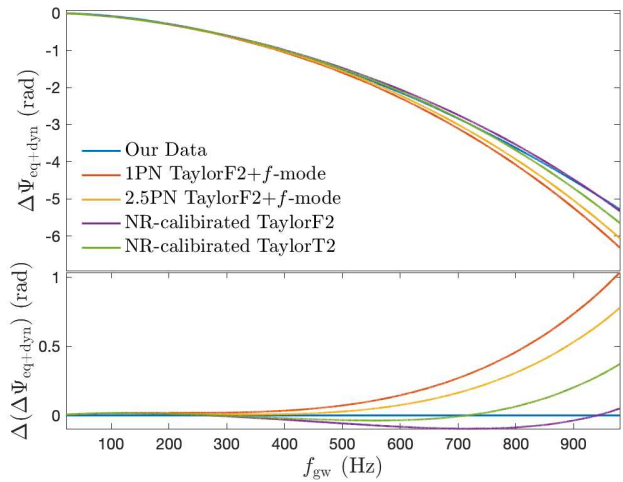}
	\caption{Tidal dephasing, including equilibrium and dynamical tides, from our code (blue) and several analytic models (see the main text) as functions of $\gw$ (top panel). The difference of each model to ours is shown in the bottom panel. For this plot, we consider a symmetric binary with $\Ms=1.3M_{\odot}=\Mc$ and the MPA1 EOS \cite{MPA1}.
	The results shown here account only for the tides in the primary, thus the total effect is twice as big.
	}
	\label{fig:comparison_psi}
\end{figure}

In Fig.~\ref{fig:comparison_Q}, we compare our result of phase acceleration $Q_{\omega}^{0}$, for a particular symmetric binary with a non-spinning primary, with ${\tilde Q}^T_\omega$ given by Eq.~\eqref{eq:PN_Q}.
We see that our result, which includes the $f$-mode effect, has a deviation $\le20\%$ from the analytic result for $\gw<500$ Hz, while larger deviation is observed for high frequencies. The growing deviation can be attributed to $f$-mode excitation, which is absent in the analytic expression \eqref{eq:PN_Q}.
To demonstrate that the deviation originates from the presence of $f$-mode, we add in the plot the phase acceleration $Q_{\omega}^{100}$, for the same binary but with the primary spinning at $100$ Hz. The enhanced $f$-mode effect in the latter spinning case manifests as the larger deviation from Eq.~\eqref{eq:PN_Q}. 
In addition, the inclusion of $f$-mode effect gives rise to a more negative phase acceleration, suggesting a faster merging (the so-called``tidally induced plunge'' \cite{Kokkotas:1995xe}).

On the other hand, in Fig.~\ref{fig:comparison_psi}, we compare the GW phase of our simulation (blue curve) with the following approaches to further demonstrate the reliability of our code: \\
\textcolor{white}{ii} (i) 1 PN TaylorF2 + $f$-mode [Eq.~\eqref{eq:1PNTaylorF2} and Eq.~\eqref{eq:fmode}], \\
\textcolor{white}{ss}(ii) 2.5 PN TaylorF2 + $f$-mode [Eq.~\eqref{eq:DTaylorF2} and Eq.~\eqref{eq:fmode}], \\
\textcolor{white}{ss}(iii) NR-calibrated TaylorF2 [Eq.~\eqref{eq:kawaguchi}], and\\ 
\textcolor{white}{ss}(iv) NR-calibrated TaylorT2 [Eq.~\eqref{eq:dietrich}].\\
Denoting the difference  between our tidal phase shift to a certain model as $\Delta(\Delta_{\text{eq+dyn}})=\Delta_{\text{eq+dyn}}^{\text{ours}}-\Delta_{\text{eq+dyn}}^{\text{model}}$, we see that the deviation from the aforementioned models is less than 1 rad overall, and most of the deviation piles up after $\gw\gtrsim400$~Hz. 
Our numerical scheme produces smaller dephasing compared to the two non-NR-corrected models, meaning the tidal effect in  our scheme is a bit weaker. However, the shifts are greater than those of NR-calibrated models when $\gw$ is less than a certain value. This ``sign-changing'' behaviour is often seen when it comes to comparing different waveform models (e.g., \cite{Lackey:2018zvw,Dietrich:2019kaq,Schmidt:2019wrl}).
Among the considered models, the model using TaylorF2 with NR waveforms is in better agreement to our result with deviation of about $\sim 0.1$ rads.

As the QNM spectrum depends on EOS and spin, we will address how they will affect the tidally-induced phase shift, notably the dependence on EOS (Sec.~\ref{UR}) and the tidal effects of spinning stars (Sec.~\ref{spin}). In general the spin itself will lead to certain dephasing due to, e.g., spin-orbit, spin-spin, and self-spin couplings. The total dephasing thus consists of the tidal and spin-included contributions. The accurate identification of the tidal part accurately is crucial in acquiring the values of the source parameters; some discussion on this issue will be provided at the end. 
Before we investigate the aforementioned aspects, we first attain confidence on the results of our code by comparing with the analytic forms obtained via PN expansions, the EOB scheme, and the phenomenological models fitting to NR simulations.

\subsection{An EOS-independent Tidal Hamiltonian}\label{UR}

In general, stellar oscillations in GR will cause perturbations in metric fields, which are damped due to GW emission in a timescale set by the imaginary part of mode frequencies. In addition, shear and bulk viscosity work to damp the mode as well. For $f$-modes, however, the viscosity damping timescale is longer than the one of gravitational damping (e.g., \cite{ipser91}).
In the present article, we will not include the damping of $f$-modes as they are gradually excited. The tidal dephasing would have been only slightly smaller if the $f$-mode damping was included.
Therefore, the small contributions from the mode-induced metric perturbation will be ignored, and the tidal parts of Hamiltonian has the form (cf.~Eqs.~(16) and (25) in \cite{Kuan:2021jmk})
\begin{align}\label{eq:tidhim}
	&H_{\text{tid}} = -\frac{2\Ms\Mc}{a\Rs}\sum_{\alpha} W_{lm}  \bigg(\frac{\Rs}{a}\bigg)^{l} 
	\Re\left[\bar{q}_{\alpha}Q_{\alpha}e^{-im\phi_{c}}\right],\\
	&H_{\text{osc}} = \frac{1}{2}\sum_{\alpha}
		\bigg( \frac{p_{\alpha}\bar{p}_{\alpha}}{\Ms\Rs^{2}}+{\Ms\Rs^{2}}\sigma_{\alpha}^{2}q_{\alpha}\bar{q}_{\alpha} \bigg) + \text{H.c.},
\end{align}
where we focus on the tidal activity in the primary. Here $\alpha$ labels different QNMs, $\phi_{c}$ is the phase coordinate of the companion, $q_{\alpha}$ are the mode amplitudes, and $p_{\alpha}$ are the canonical momenta associated to $q_{\alpha}$. The (inertial-frame) eigenfrequency of the excited mode is $\sigma_{\alpha}=\omega_{\alpha}+i/\tau_{\alpha}$, where $\omega_{\alpha}$ is the frequency, and $\tau_{\alpha}$ the damping timescale. The overhead bar denotes complex conjugation. In the present article, we investigate the tidal excitation of modes in NSs with \emph{aligned} spins, thus only $l=m$ modes will be relevant. In addition, as stated above, we will limit our study to the $l=2=m$ $f$-mode hence we will drop the subscript $\alpha$ hereafter, and denote its coupling strength as $Q_{f}$, which should not be confused with the phase acceleration $Q_{\omega}$ defined in Eq.~\eqref{eq:phiacc}.

\begin{figure}
	\centering
	\includegraphics[width=\columnwidth]{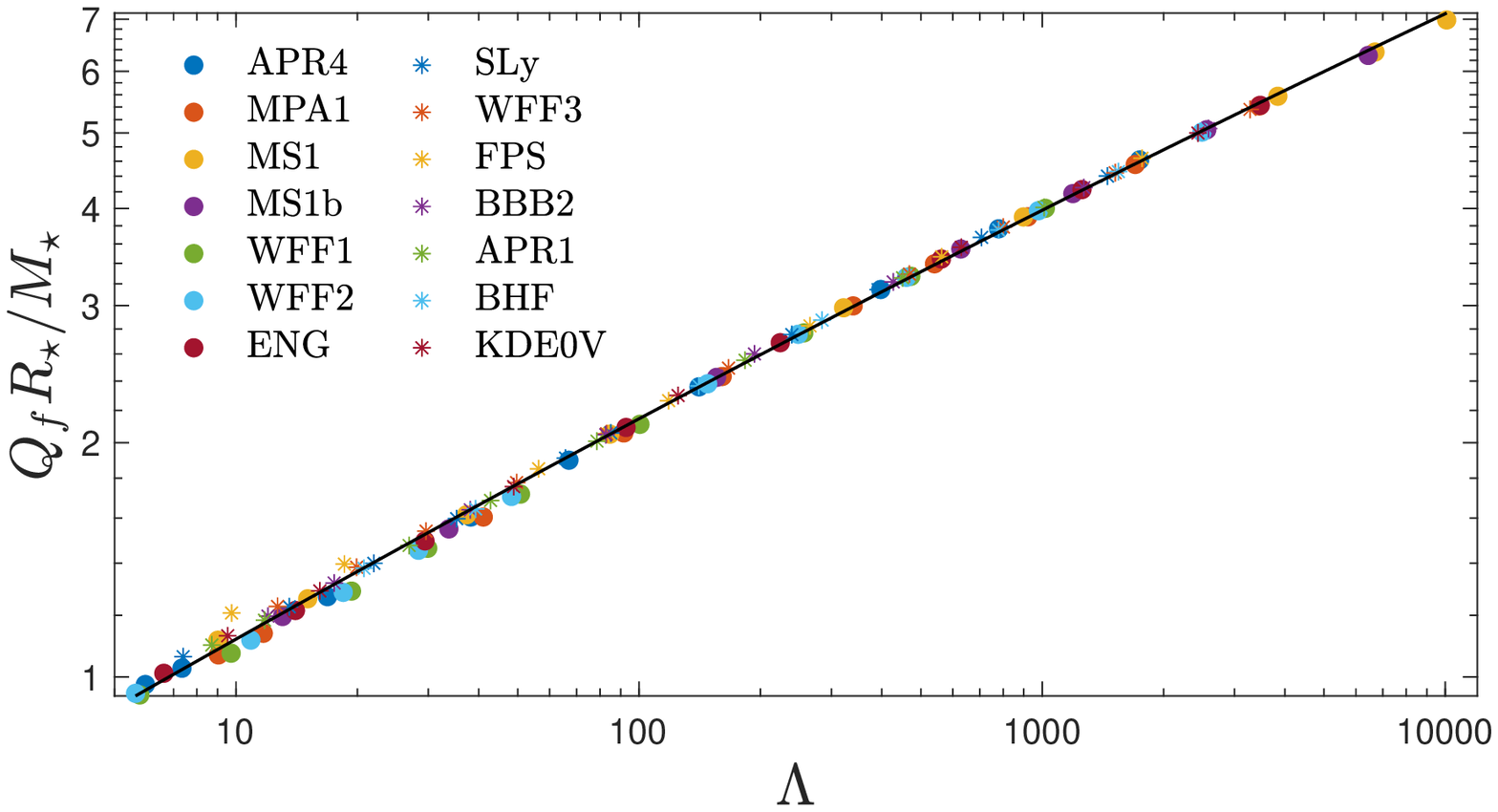}
	\includegraphics[width=\columnwidth]{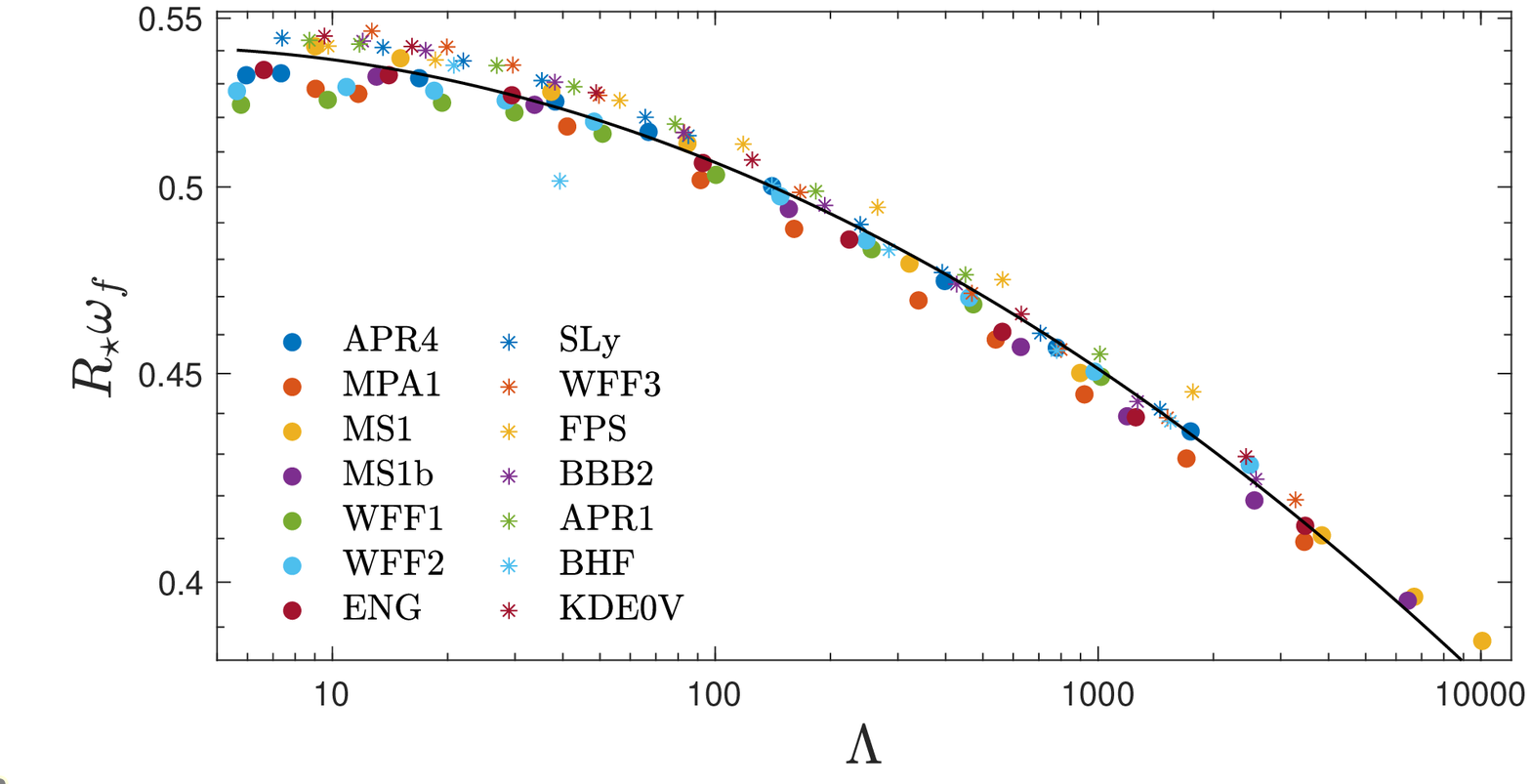}
	\caption{Universal relations connecting $Q_f\Rs/\Ms$ [Eq.~\eqref{eq:uniB}; top panel] and $\Rs^2\omega_f^2$ [Eq.~\eqref{eq:uniA}; bottom panel] to the tidal deformability of the primary $\Ls$. 
	}
	\label{fig:uni}
\end{figure}

We introduce the primary-based, dimensionless quantities
\begin{subequations}
\begin{align}
	&{\cal A}(\Ls)=Q_f\Rs/\Ms,\\
	&{\cal B}(\Ls)=\Rs\omega_f,
\end{align}
\label{eq:Uni_AB}
\end{subequations}
which, if the primary is non-spinning (i.e., $\omega_f$ is the mode frequency), can be expressed as functions of $\Lambda$. From the numerical values of ${\cal A}$ and ${\cal B}$, we find the following fitting
\begin{align}
	\log {\cal A} = -0.2887 + 0.2766 \log\Ls -0.0094 (\log\Ls)^2,\label{eq:uniA}
\end{align}	
and
\begin{align}
	\log {\cal B} = -1.0644 + 0.0001 \log\Ls -0.0096 (\log\Ls)^2.\label{eq:uniB}
\end{align}
The relations are plotted in Fig.~\ref{fig:uni}, where the considered set of EOS is labeled in the legend. This set of EOS is the same as those adopted in \cite{Kuan:2022bhu}, and we note that the latter formula \eqref{eq:uniB} has been introduced there. The former universal relation connects the tidal overlap of $f$-mode to the tidal deformability, which is presented for the first time here, while that for $r$-mode has been developed in \cite{Ma:2020oni} (see the right panel of Fig.~7 therein).

Substituting the previous quantities into the tidal Hamiltonian, we get
\begin{align}\label{eq:uni_H}
	H^{T}=H_{\text{tid}}+H_{\text{osc}}=&-\frac{2\Ms^2\Mc}{a^3} {\cal A} q\cos(m\varphi_c) \nonumber\\
	&+\frac{p\bar{p}}{\Ms\Rs^2}+\Ms q \bar{q} {\cal B}^2.
\end{align}
which is a functional depending on the individual masses $\Ms$ and $\Mc$, the tidal deformability of the primary $\Ls$, and $\Ms\Rs^2$. The latter quantity is related to moment of inertia \cite{Lattimer:2004nj} [see Eq.~(12) therein].
The dependencies are all detectable in GW analysis either directly or indirectly;  the measurement of the chirp mass $\mathcal{M}$ and the symmetric mass ratio $\eta$ determine the individual masses, while the measurement of $\tilde{\Lambda}$ returns $\Rs$ in a manner independent of the $\Mc$ if the chirp mass is known \cite{Raithel:2018ncd,Zhao:2018nyf}. In addition, the mass ratio together with $\tilde{\Lambda}$ estimate the individual tidal deformabilities since $\Ls$ and $\Lc$ relate to each other by (see Eq.~8 of \cite{De:2018uhw})
\begin{align}
	\Ls \simeq q^6 \Lc.
\end{align} 
The tidal Hamiltonian can therefore be EOS-independently reconstructed from $\mathcal{M}$, $\eta$, and $\tilde{\Lambda}$.
\emph{To our knowledge, the universality of the tidal phase shift, especially the contributions of dynamical tides, has not yet been recognised in the literature. } 

\begin{figure}
	\centering
	\includegraphics[width=\columnwidth]{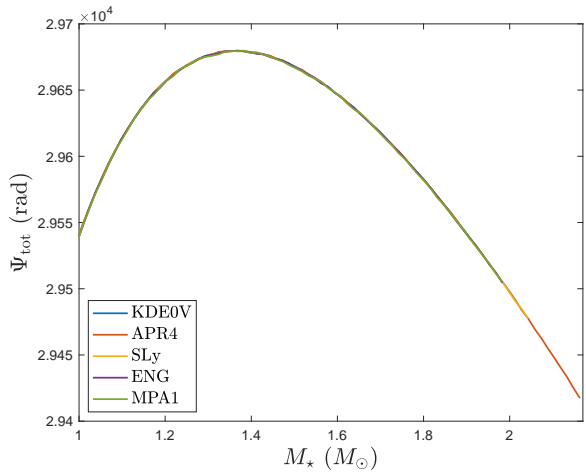}
	\caption{$\Psi_{\text{tot}}$ [Eq.~\eqref{eq:phiT}] for binaries with a fixed chirp mass $\mathcal{M}=1.186M_{\odot}$ as functions of $\Ms$ for the chosen EOS. Each curve terminates at the maximal mass of the associated EOS.
	}
	\label{fig:const_chirp} 
\end{figure}

The significance of these three parameters in determining GW phasing goes, descendingly, from the chirp mass to the tidal deformability.
Denoting the GW phase accumulated when $\gw$ ranging between $f_{\text{min}}=20$ Hz and $f_{\text{max}}=1000$ Hz as
\begin{align}\label{eq:phiT}
	 \Psi_{\text{tot}} &= \int_{f_{\text{min}}}^{f_{\text{max}}} d\gw\bigg(\frac{\partial \Psi}{\partial \gw}\bigg){}_{\Ls,\Lc,\nus,\nu_{\text{s,comp}}}\nonumber\\
	 &=\Psi_{\text{pp}} + \Delta\Psi^T,
\end{align}
we plot for five EOS in Fig.~\ref{fig:const_chirp} the accumulated GW phase for binaries with fixed chirp mass $\mathcal{M}=1.186M_{\odot}$ as functions of $\Ms$.
Here $\Psi_{\text{pp}}$ is the part of point-particle contribution, and $\Delta\Psi^T$ is the tidal dephasing due to both equilibrium and dynamical tides. The mass ratio of considered binaries can be obtained from $\mathcal{M}$ and $\Ms$. We note that the chosen EOS span a wide range of stiffness going from the stiffest MPA1 down to the softest KDE0V. 
The binaries considered in Fig.~\ref{fig:const_chirp} undergo $\sim2260$ orbits in the last $\sim150$ s of the coalescence, during which $\gw$ climbs from 20 Hz to $10^3\text{ Hz}$. This corresponds to $\sim4560$ cycles of \emph{time-domain} gravitational waveform, while the \emph{frequency-domain} gravitational waveform \eqref{eq:SPA} is found to oscillate $\sim4775$ cycles, i.e., $\Psi_{\text{tot}}\lesssim3\times10^4$ rads; we recall that the time- and frequency-domain phasing of waveforms are connected via the relation \eqref{eq:psi_lagendre}.
The phase varies $\lesssim1\%$ for different $\eta$, while the variance of the finite size effects encoded in $\tilde{\Lambda}$ is even smaller.
In addition, the phase peaks at $\Ms=\Mc \simeq 1.37M_{\odot}$, indicating that symmetric binaries will undergo more cycles before merger.

\subsection{Spin modification in Tidal Dephasing}\label{spin}

In spinning neutron stars the oscillation frequency splits into co- and counter-rotational components.
The inertial-frame $f$-mode frequency, $\omega_f$, of the primary shifts according to (cf.~Eq.~(70) in \cite{Kuan:2021jmk})
	\begin{align}\label{eq:modrot}
		\delta\omega^{R}_{f} = -2\pi m (1-C_{f})\nus,
	\end{align}
where $C_{f}$ depends on  the eigenfunction of $f$-mode via the integration of Eq.~(71) in \cite{Kuan:2021jmk}. 
For a fast-rotating primary, the frequency shift involving  extra terms quadratic in $\nus$ has been proposed in \cite{Kruger:2019zuz, 2020PhRvD.102f4026K, 2021FrASS...8..166K}. 
This quadratic term is however negligible up to a spin of $\lesssim$ 1kHz (cf.~Fig.~1 of \cite{2021FrASS...8..166K}), about half the non-rotating mode frequency, thus the linear modification shown in Eq.~\eqref{eq:modrot} is adequate for our purpose.
This shift \eqref{eq:modrot} is negative for the $l=2=m$ $f$-mode given that $C_f\approx 0.3$, leading to a smaller oscillation frequency. As a consequence, the resonance between the mode and the orbital frequency occurs earlier, rendering stronger tidal dephasing. We note that the strength of the tidal dephasing depends on the alignment of the spin in general. In addition, modes with $m\ne2$ will be excited as well in misaligned stars. 
The influence of the tilt angle of spin in the tidal dephasing will be deferred to future study.

\begin{figure}[h!]
	\centering
	\includegraphics[width=\columnwidth]{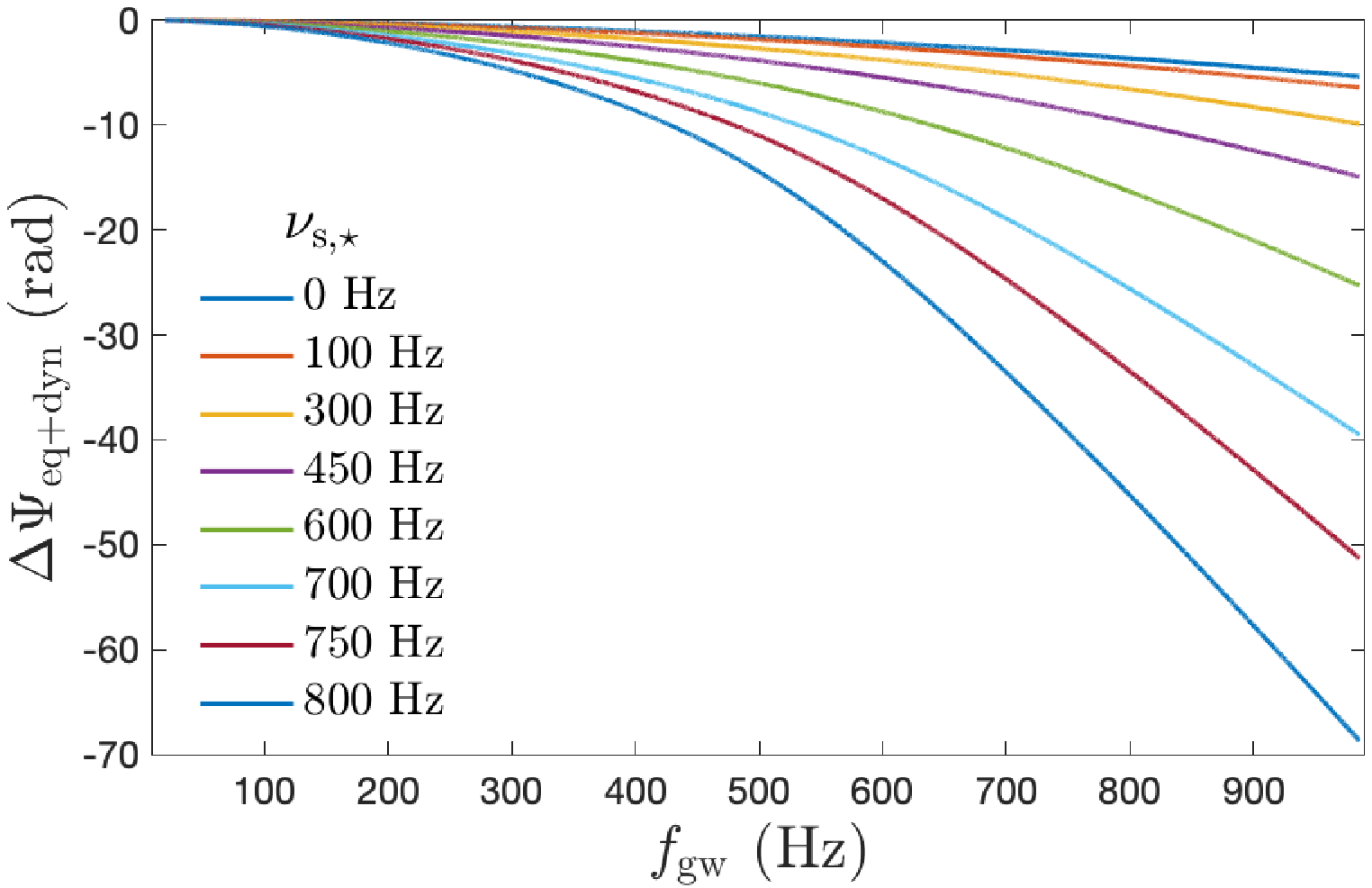}
	\includegraphics[width=\columnwidth]{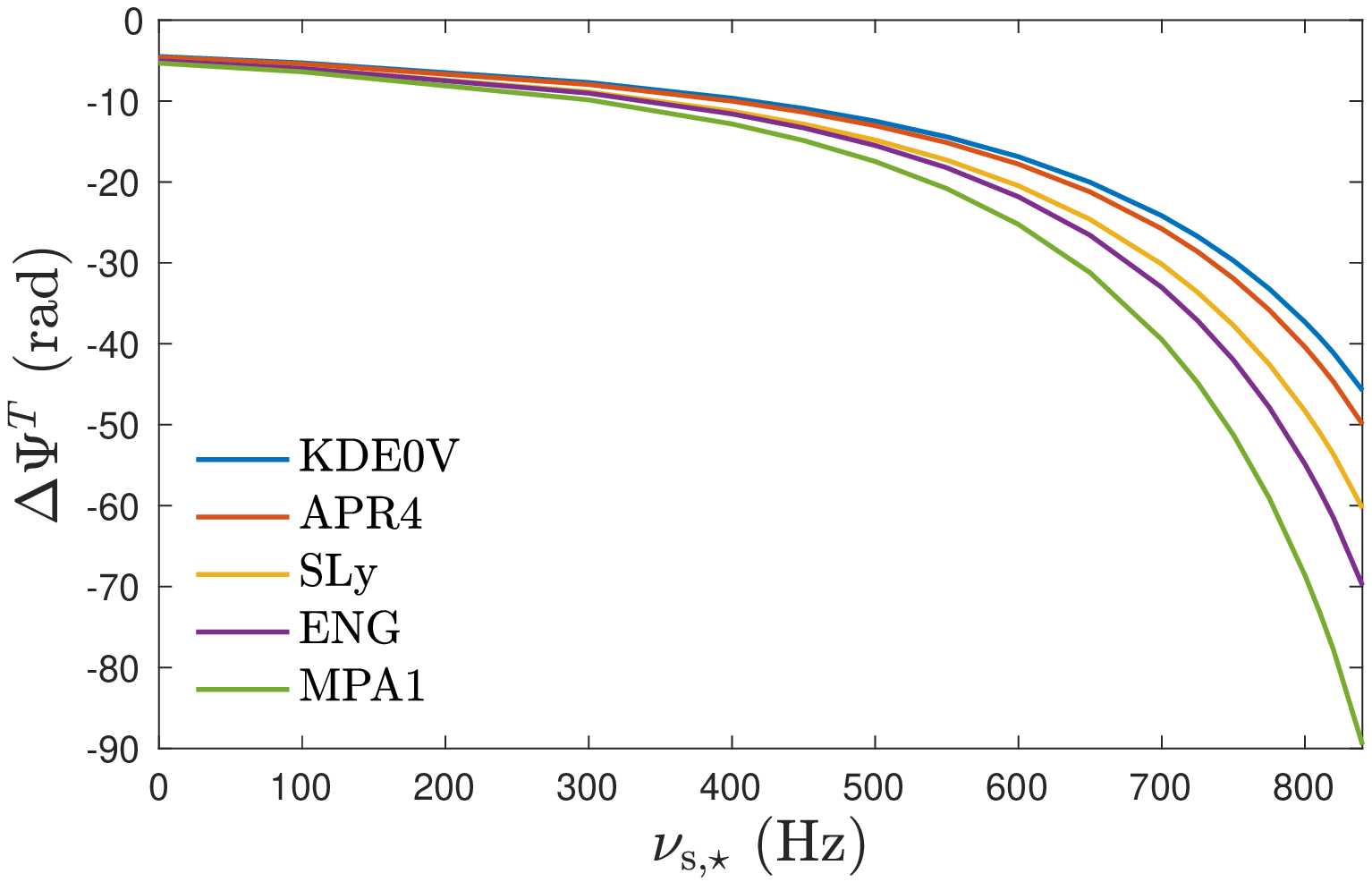}
	\caption{\emph{Top panel}:  Evolution of GW phase for various spin rates of the primary, $\nus$, as functions of $\gw$ for a particular binary (we assumed the MPA1 EOS).
	\emph{Bottom panel}: Accumulated phase shift $\Delta\Psi^T$ due to tidal effects in both NSs as a function of $\nus$, 	here we used the same EOS  as in Fig.~\ref{fig:const_chirp}. For both panels, we consider the tidal effect of the primary in symmetric binaries with a fixed tidal deformability $\tilde{\Lambda}=920$.
	}
	\label{fig:spin}
\end{figure}

For symmetric binaries with $\tilde{\Lambda}=920$, and two NSs spinning at the same rate, we plot in the top panel of Fig.~\ref{fig:spin} the tidal dephasing induced by the primary for various spins,  as functions of $\gw$. 
We see that the phase shift increases monotonically with $\nus$, while a noticeable, rapid growth is observed in the high GW frequency regime.
Taking the case with $\nus=800\text{ Hz}$ for instance, the dephasing piles up to about $20$ rads as  $\gw$ goes from 20 Hz to 500 Hz, while the accumulated dephasing is  $\lesssim 150$ rads from $\gw=500$ to $10^3\text{ Hz}$. 
In the bottom panel of Fig.~\ref{fig:spin}, we show the tidal dephasing $\Delta\Psi^T$ defined in Eq.~\eqref{eq:phiT} as  function of $\nus$ for the five EOS used in Fig.~\ref{fig:const_chirp}. We observe that the waveform dephasing depends on the EOS, and is smaller for softer EOS due to the smaller radius of the star or equivalently the higher compactness . 
Again, we witness a rapid increase of $|\Delta\Psi^T|$ for higher spin due to a different reason: For cases with fixed spin, the dephasing grows faster for a cutoff $f_{\text{max}}>500$ Hz since the information of dynamical tides lies in the high frequency part of waveforms. Moreover when a fixed $f_{\text{max}}$ is assumed, the dephasing enlarges due to earlier excitation of $f$-mode.
Given that the dephasing caused by the equilibrium tide is only $\lesssim10$ rads, the contribution of $f$-mode excitation is dominant for $\nus>200$~Hz. In particular, the dephasing of $\Delta\Psi^T\sim 90$ rads at the right end of the green curve in the bottom panel of Fig.~\ref{fig:spin} is mainly caused by $f$-mode excitation.

Having seen that the Hamiltonian \eqref{eq:uni_H} for binaries consisting of non-spinning stars can be EOS-independently written down, we further find that this universality survives even for small stellar spins $\lesssim 100$ Hz.
For example, the phase shift $\Delta\Psi^T$ due to the $f$-mode in the primary is EOS-independent for $\nu_{\text{s},\star}\lesssim100$ Hz, as shown in the bottom panel of Fig.~\ref{fig:spin}.
The reason is as follow: In reality, mode frequency will be modified differently depending on the EOS. Although this difference is minor for small spins, the dependence of the phase shift on the EOS is becoming noticeable as the stellar spin increases.
The Hamiltonian is EOS-independent for most of the coalescing NS binaries since its members are typically old and tend to rotate slowly. Nevertheless, there may still be a number of  NSs in binaries with moderate and higher spin rates. A potential case is the secondary of GW190814 \cite{Hannam:2013uu}, we will discuss the specific case in Sec.~\ref{case0814}.

\begin{figure}
	\centering
	\includegraphics[width=\columnwidth]{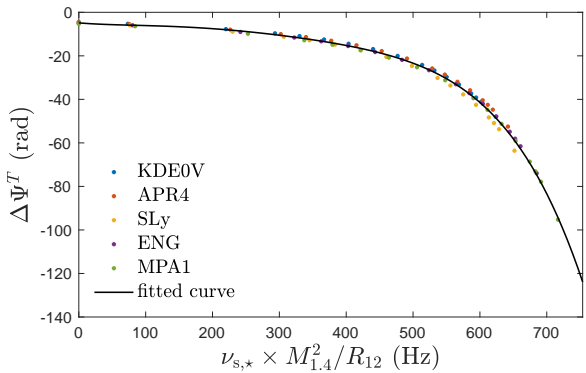}
	\caption{Tidal dephasing as a universal function of the dimensionless, normalised spin defined in Eq.~\eqref{eq:dimless_norm_spin}. Here we denote $M_{1.4}=\Ms/1.4 M_{\odot}$ and $R_{12}=\Rs/12$ km.
	}
	\label{fig:uni_psi_spin}
\end{figure}

In addition, we find that $\Delta\Psi^T$ can be numerically EOS-insensitively parameterised by a normalised spin, defined by
\begin{align}\label{eq:dimless_norm_spin}
    \tilde{\nu} = \nus\left( \frac{\Ms}{1.4M_{\odot}} \right)^2\left( \frac{\Rs}{12\text{ km}} \right),
\end{align}
for a fixed $\tilde{\Lambda}$. For example, a universal relation,
\begin{align}\label{eq:uni_psi_spin}
    \Delta\Psi^T=& -4.850-2.539\times10^{-2}\tilde{\nu}+2.449\times10^{-4}\tilde{\nu}^2\nonumber\\
    &-1.429\times10^{-6}\tilde{\nu}^3+3.026\times10^{-9}\tilde{\nu}^4\nonumber\\
    &-2.482\times10^{-12}\tilde{\nu}^5 \text{ rad}
\end{align}
is found for $\tilde{\Lambda}=920$.
The relation \eqref{eq:uni_psi_spin} is plotted in Fig.~\ref{fig:uni_psi_spin}.

Apart from the rotation-induced modifications to the $f$-mode frequency $\omega_f$, rotation will induce a correction $\delta Q_f$ to the tidal overlap integral $Q_f$. This term, however, is of  quadratic order to $\nus$ \cite{Landry:2015zfa,Pani:2015hfa,pnig22}, and is typically $\delta Q_f\lesssim 10^{-3} Q_f$ even for the fastest rotating NS in a binary observed to-date, viz.~PSR J0737-3039A, whose dimensionless spin is $\chi\lesssim0.05$ ($\sim44$ Hz; \cite{Burgay:2003jj}).
However, the secondary of GW190814 has a finite chance to set a new record with $\chi\sim0.47$ ($\sim1170$ Hz \cite{Biswas:2020xna}; see also below). In the latter system, a correction of $\delta Q_f\gtrsim0.2 Q_f$ is expected. However, we will not explore this interesting case  in the present article, and we will base our estimations on the tidal overlap integral for non-rotating stars.

\subsection{Statistical Error}\label{stat.err}

The phase of the (frequency-domain) GW waveform is particularly crucial in estimating parameters in the matching filter algorithm \cite{Cutler:1992tc,Cutler:1994ys,Finn:1992xs}, which we summarise in the following. 

Defining a sensitivity-curve-weighted inner product in the waveform-space as 
\begin{align}\label{eq:inner}
	(g|h) = 2\int_{0}^{\infty}\frac{g^*(f)h(f)+h(f)^*g(f)}{S_n(f)}df,
\end{align}
for two frequency-domain waveforms $f$ and $g$, the SNR can then be express as
\begin{align}
	\rho^2[h] = (h|h)=4\int_0^{\infty}\frac{|h(f)|^2}{S_n(f)}df=4\mathcal{A}^2\int_0^{\infty}\frac{f^{-7/3}}{S_n(f)}df,
\end{align}
where $h$ is the input waveform template, and the latter equality holds if SPA \eqref{eq:SPA} has been adopted. Here $S_{n}(f)$ is the sensitivity curve set by the detector, and the superscript asterisk denotes complex conjugate.

As any measurement comes along with errors, we have to handle the posterior possibility of getting a somewhat different set of parameters $\mathbf{\theta}$, which deviates from the true parameters  $\mathbf{\theta}_o$, by a minute inaccuracy $\Delta\mathbf{\theta}_o$ for a given signal $s$, i.e., $p(\mathbf{\theta}|s)$ must be under control. For a large S/N, the approximation for the posterior possibility,
\begin{align}
	p(\mathbf{\theta}|s)\propto\exp\big[-\frac{1}{2}\Gamma_{ab}\Delta\theta^a\Delta\theta^b \big],
\end{align}
exhibits a Gaussian distribution around $\mathbf{\theta}_o$ \cite{Poisson:1995ef,Cutler:1994ys}, which is characterised by the Fisher information matrix
\begin{align}\label{eq:fisher}
	\Gamma_{ab} = \bigg(\frac{\partial h}{\partial \theta^a}\bigg\vert\frac{\partial h}{\partial \theta^b}\bigg).
\end{align}
The measurement error of $\theta^a$ is then defined as
\begin{align}\label{eq:error}
	\sigma_a=\sqrt{(\Gamma^{-1})^{aa}},
\end{align}
where $\Gamma^{-1}$ is the inverse of the Fisher matrix.

Neglecting explicitly spin-related terms, which is appropriate when the spin is well-constrained (see discussion) or rather small\footnote{As of the time this article is prepared, the known, fastest spinning NS in binaries is PSR J0737-3039A, whose dimensionless spin, though depending on the EOS, is estimated to be $\lesssim0.03$ \cite{Damour:2012yf}.} \cite{Damour:2012yf}, we have the symbolic expression
\begin{align}\label{eq:strain}
	h=h(\gw;\mathcal{A},f_0t_o,\phi_o,\mathcal{M},\eta,\tilde{\Lambda},\omega_f)
\end{align}
focusing on the explicit dependencies. Here $f_0$ is the frequency at the minimum of the sensitivity curve, and we recall that $t_o$ and $\phi_o$ are the reference time and phase often set as the values at the merger [defined in Eq.~\eqref{eq:psi_lagendre}]. Although the Fisher matrix is 7 dimensional, we can suppress one of its dimensions by factoring out the elements related to  $\mathcal{A}$ since the amplitude is uncorrelated with the other quantities involved in the inner product Eq.~\eqref{eq:inner} in SPA.

\begin{table*}
	\centering
	\caption{Statistical estimation of the  measurement error for the accumulated tidal dephasing $\Delta\Psi^T$  [Eq.~\eqref{eq:phiT}] (3rd column), the GW phase $\Delta\phi_o$ (4th column), the chirp mass $\Delta\mathcal{M}/\mathcal{M}$ (5th column), the symmetric mass ratio $\Delta\eta/\eta$ (6th column), the mutual tidal deformability  $\Delta\tilde{\Lambda}/\tilde{\Lambda}$ (7th column), and the frequency of the $f$-mode in the primary $\Delta\omega_{f}/\omega_{f}$ (8th column) assuming the cutoff as $f_{\text{max}}=10^3$ Hz. We additionally considered the uncertainty of $\tilde{\Lambda}$ for $f_{\text{max}}=450$ Hz in the final column so as to be compared to the results in \cite{Damour:2012yf}, which are shown in the final column of their Tab.~2 though there the authors adopted $\rho=1$ and did not present by percentage.
	We consider symmetric binaries with $\tilde{\Lambda}=920$ for several EOS listed in the 1st column. We use four representative spin rates (given in Hz): 0 (non-spinning), 45 (fastest known NS in binaries), 500 (moderate fast), and 800 (rather fast) of the primary. 
  In general, the errors scale as $1/\rho$, while the results derived  assuming  an SNR value of $\rho=25$. The table are prepared by considering only the tidal effects of the primary.}
	\begin{ruledtabular}
	\begin{tabular}{cccccccc|c}
		EOS & $\nus$ (Hz) & $\Delta\Psi^T$ (rad) & $\Delta\phi_o$ (rad) & $\Delta\mathcal{M}/\mathcal{M}$ & $\Delta\eta/\eta$ & $\Delta\tilde{\Lambda}/\tilde{\Lambda}$ & $\Delta\omega_{f}/\omega_{f}$ & $[\Delta\tilde{\Lambda}/\tilde{\Lambda}]_{450}$ \\
		\hline 
		KDE0V & 0 & -4.4982 & $1.5231$ & $0.0037\%$ & $1.1383\%$  & $5.0868\%$ & $664.5\%$ & 83.67$\%$ \\
		& 45 & -4.8299 & $1.5150$ & $0.0037\%$ & $1.1335\%$  & $4.4050\%$ & $568.3\%$ & 73.23$\%$ \\
		& 500 & -12.4849 & $1.6379$ & $0.0038\%$ & $1.2128\%$  & $0.6339\%$ & $74.08\%$ & 5.940$\%$ \\
		& 800 & -37.3150 & $1.4969$ & $0.0037\%$ & $1.1274\%$  & $0.2266\%$ & $14.92\%$ & 2.080$\%$ \\
		\hline
		APR4 & 0 & -4.5928 & $1.5220$ & $0.0038\%$ & $1.1496\%$  & $4.8817\%$ & $634.9\%$ & 79.62$\%$ \\
		& 45 & -4.9372 & $1.5138$ & $0.0038\%$ & $1.1447\%$  & $4.2180\%$ & $541.5\%$ & 69.74$\%$ \\
		& 500 & -13.0437 & $1.4196$ & $0.0037\%$ & $1.0886\%$  & $0.7189\%$ & $70.23\%$ & 11.99$\%$ \\
		& 800 & -40.4013 & $1.5165$ & $0.0038\%$ & $1.1514\%$  & $0.2151\%$ & $13.79\%$ & 1.777$\%$\\
		\hline 
		SLy & 0 & -5.0406 & $1.5218$ & $0.0039\%$ & $1.1630\%$  & $4.3130\%$ & $557.2\%$ & 69.71$\%$ \\
		& 45 & -5.4287 & $1.5137$ & $0.0039\%$ & $1.1581\%$  & $3.7168\%$ & $473.6\%$ & 60.76$\%$ \\
		& 500 & -14.8164 & $1.4367$ & $0.0038\%$ & $1.1124\%$  & $0.6322\%$ & $60.40\%$ & 9.857$\%$ \\
		& 800 & -48.2919 & $1.5813$ & $0.0040\%$ & $1.2057\%$  & $0.1956\%$ & $12.18\%$ & 1.300$\%$ \\
		%
		%
		\hline 
		ENG & 0 & -5.0167 & $1.5184$ & $0.0042\%$ & $1.1828\%$  & $4.1548\%$ & $531.8\%$ & 66.64$\%$ \\
		& 45 & -5.4175 & $1.5110$ & $0.0042\%$ & $1.1782\%$  & $3.5738\%$ & $450.8\%$ & 57.58$\%$\\
		& 500 & -15.4857 & $1.4449$ & $0.0041\%$ & $1.1388\%$  & $0.5933\%$ & $54.78\%$ & 8.660$\%$ \\
		& 800 & -54.8652 & $1.6344$ & $0.0044\%$ & $1.2623\%$  & $0.1858\%$ & $10.93\%$ & 0.9655$\%$ \\
		\hline 
		MPA1 & 0 & -5.3222 & $1.5179$ & $0.0045\%$ & $1.2083\%$  & $3.7394\%$ & $473.2\%$ & 58.69$\%$ \\
		& 45 & -5.7651 & $1.5102$ & $0.0045\%$ & $1.2034\%$  & $3.2010\%$ & $398.6\%$ & 49.91$\%$ \\
		& 500 & -17.4710 & $1.2997$ & $0.0042\%$ & $1.0679\%$  & $0.6222\%$ & $17.48\%$ & 14.19$\%$ \\
		& 800 & -68.6082 & $1.7409$ & $0.0049\%$ & $1.3587\%$  & $0.1700\%$ & $9.410\%$ & 0.6029 $\%$ \\
	\end{tabular}
	\end{ruledtabular}
	\label{tab:snr}
\end{table*}

For the measurement of $\mathcal{M}$ and $\eta$, it has been demonstrated in \cite{Damour:2012yf} that 2 PN order approximants for point-mass waveform suffice the purpose of estimating errors in tidal parameters. For later convenience, we provide the derivatives (cf.~Eq.~(3.10) of \cite{Poisson:1995ef})
\begin{widetext}
\begin{subequations}\label{eq:dh_pp}
\begin{align}
	\frac{\partial \ln h_{\text{pp}}}{\partial (f_0t)}=&2\pi i(\gw/f_0),\\
	\frac{\partial \ln h_{\text{pp}}}{\partial\phi_o}=&-i,\\
	\frac{\partial \ln h_{\text{pp}}}{\partial \ln\mathcal{M}}=&-\frac{5i}{128}(\pi\mathcal{M}\gw)^{-5/3}\bigg[1
		+\bigg(\frac{743}{252}+\frac{11}{3}\eta\bigg)x-\frac{32\pi}{5}x^{3/2}+\bigg(\frac{3058673}{508032}+\frac{5429}{504}\eta+\frac{617}{72}\eta^2\bigg)x^2\bigg],\\
\intertext{and}
	\frac{\partial \ln h_{\text{pp}}}{\partial \ln\mathcal{\eta}}=&-\frac{i}{96}(\pi\mathcal{M}\gw)^{-5/3}\bigg[
		\bigg(\frac{743}{168}-\frac{33}{4}\eta\bigg)x
		-\frac{108}{5}\pi x^{3/2}+\bigg(\frac{3058673}{56448}-\frac{5429}{224}\eta-\frac{5553}{48}\eta^2\bigg)x^2\bigg]
\end{align}
\end{subequations}
\end{widetext}
for the point-mass approximants. We will approximate the variance of the strain \eqref{eq:strain} with infinitesimal changes of non-tidal parameters by the point-mass formulas \eqref{eq:dh_pp}, i.e., 
\begin{align}
    \frac{\partial h}{\partial X}\simeq\frac{\partial h_{\text{pp}}}{\partial X}
\end{align}
for $X=\{f_0t, \, \phi_o, \, \ln\mathcal{M}, \,\ln\eta\}$.
On the other hand, we numerically evaluate $\frac{\partial h}{\partial \tilde{\Lambda}}$ by first constructing two waveforms with slightly different tidal deformabilities $\tilde{\Lambda}\pm\epsilon$, while fixing other parameters, then dividing the difference of the two waveforms by the difference in $\tilde{\Lambda}$, viz.
\begin{align}
    \frac{\partial h}{\partial \tilde{\Lambda}}=\frac{h(\tilde{\Lambda}+\epsilon)-h(\tilde{\Lambda}-\epsilon)}{2\epsilon}.
\end{align}
The scheme admits that the variation of $h(\Lambda)$ does not feel the variance of the others (even though via the inversion of the Fisher matrix there is some mixing).
The same procedure is performed for the parameter $\omega_f$ to obtain $\partial h/\partial \omega_f$. Measurement errors, \eqref{eq:error}, can then be calculated by inverting the Fisher matrix \eqref{eq:fisher}.

Adopting the sensitivity curve of aLIGO and assuming that the data streams are measured across the frequency band $20\le\gw\le10^{3}$ Hz with a SNR $\rho=25$, we estimate the errors $\Delta\mathcal{M}/\mathcal{M}$,  $\Delta\eta/\eta$, $\Delta\tilde{\Lambda}/\tilde{\Lambda}$ and $\Delta\omega_{f}/\omega_{f}$ in Table~\ref{tab:snr}. We see that the magnitude of tidal phase shift $\Delta\Psi^T$ increases with stellar spin due to earlier excitation of $f$-mode, allowing for a more accurate extraction of tidal parameters. In particular, the error in $\tilde{\Lambda}$ and $\omega_f$ reduce rapidly for increasing spin, where the error can be $<1\%$ for the former and $<15\%$ for the latter if the NS spins at 800 Hz.
The improvement in the measurability of both $\tilde{\Lambda}$ and $\omega_f$ is due to the earlier excitation of the $f$-mode whose frequency was shifted by the rotation. An earlier mode excitation increases significantly the transfer of orbital energy to stellar oscillations affecting significantly the dephasing.
As a result, $f$-mode frequency will be estimated with significantly smaller error.
Actually, even thought the dephasing due to the equilibrium tide is not directly affected, the increasing influence of the dynamical tides encodes certain information of the equilibrium tides since the latter is the adiabatic limit of the former -- notice that $\Lambda$ factors out  in $\Delta\Psi^{\text{dyn}}_{\star}$ in Eq.~\eqref{eq:fmode}. Therefore, by including the dynamical tides in the Fisher analysis for $\tilde{\Lambda}$ we effectively place extra emphasis on the high-frequency part of waveform.

\begin{figure}
	\centering
	\includegraphics[width=\columnwidth]{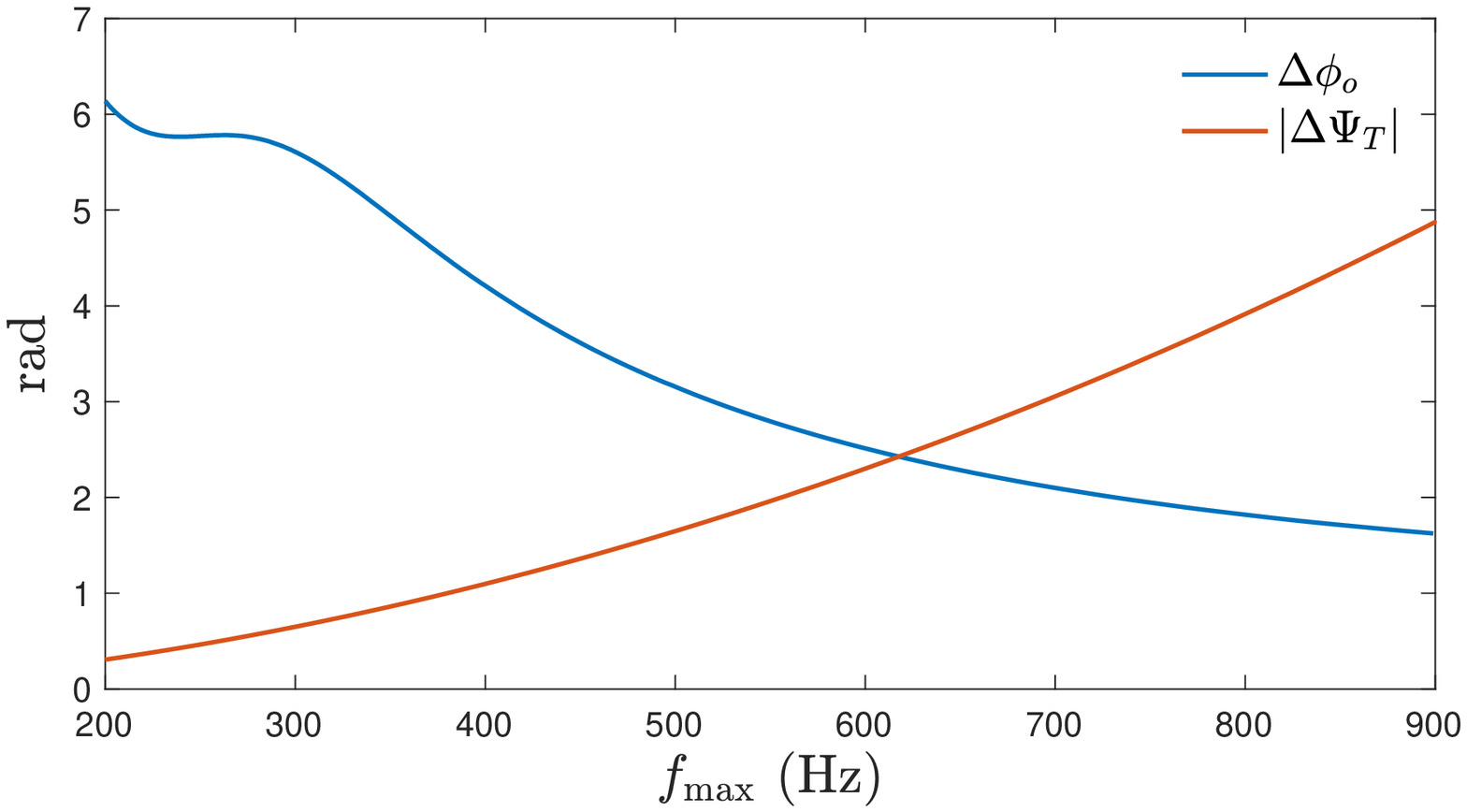}
	\includegraphics[width=\columnwidth]{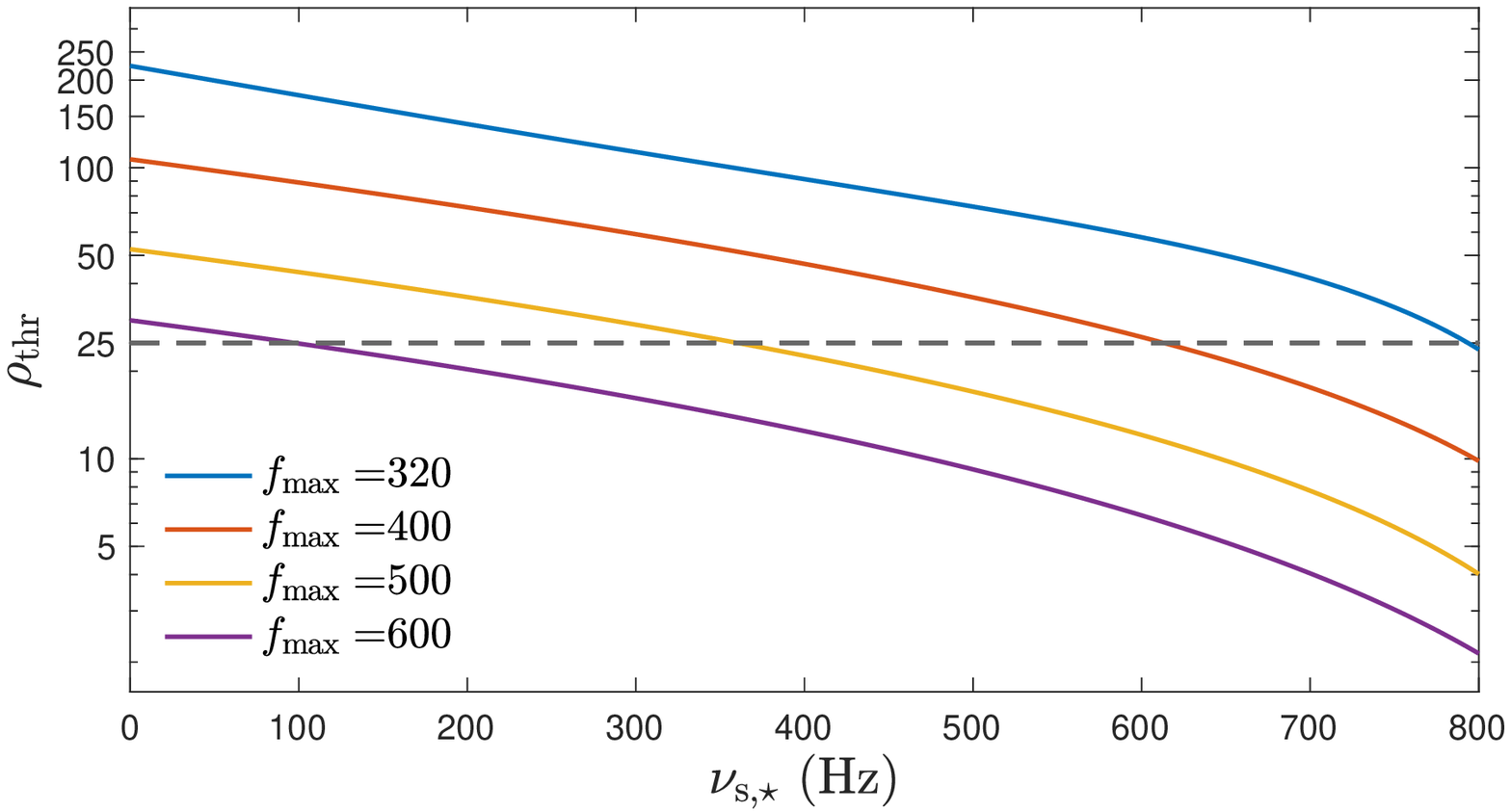}
	\caption{
	\emph{Top panel}: Uncertainty in the GW phase $\Delta\phi_o$ (blue) and tidally-induced phase shift {$\Delta\Psi^T$} (red) [Eq.~\eqref{eq:phiT}] as functions of the cutoff frequency $f_{\text{max}}$. The spin of the primary and the SNR are set as, respectively, $\nus=45$ Hz and $\rho=25$. 
	\emph{Bottom panel}: SNR $\rho_{\text{thr}}$ for the uncertainty in phase $\Delta\phi_o$ to equate the tidal dephasing $|\Delta\Psi^T|$ [Eq.~\eqref{eq:snr_thr}] for four cutoff frequencies as functions of $\nus$. In both panels, the symmetric binary with $\tilde{\Lambda}=920$ and EOS MPA1 is considered.
	} 
	\label{fig:compare}
\end{figure}

For the considered data stream, the tidal dephasing $\Delta\Psi^T$ is larger than the uncertainty of phase $\Delta\phi_o$ even for a non-spinning star. However, the tidal dephasing may be hidden in the uncertainty in phase $\Delta\phi_o$ for a lower cutoff.
In general, tidal dephasing is a function of $\nus$ and $f_{\text{max}}$ [i.e., $\Delta\Psi_T=\Delta\Psi_T(\nus,f_{\text{max}})$], while the uncertainty in phase is a function of SNR and $f_{\text{max}}$ [i.e., $\Delta\phi_o=\Delta\phi_o(\rho,f_{\text{max}})$].
For a particular binary with 
the spin of the primary being $\nus=45\text{ Hz}$ and the SNR of the associated waveform being $\rho=25$, we integrate Eq.~\eqref{eq:phiT} and Eq.~\eqref{eq:inner} from $\gw=20\text{ Hz}$ to a varying cutoff $f_{\text{max}}$.
We plot $\Delta\phi_o$ and $\Delta\Psi^T$ as functions of $f_{\text{max}}$ in the top panel of Fig.~\ref{fig:compare}. 
In this example, the tidal dephasing becomes noticeable if the cutoff is $\gtrsim600\text{ Hz}$. 
In general setting, there is a minimal SNR $\rho_{\text{thr}}$ for a specific spin $\nu_o$ and cutoff $f_{\text{max},o}$, defined by the equality
\begin{align}\label{eq:snr_thr}
    \Delta\Psi^T(\nu_o,f_{\text{max,o}})=\Delta\phi_o(\rho_{\text{thr}},f_{\text{max},o}).
\end{align}
To grasp how the increasing spin improves the detectability of tidal effects, we find $\rho_{\text{thr}}$ as function of the NS spin assuming some cutoff frequencies for a particular binary in the bottom panel of Fig.~\ref{fig:compare}. Improvement of measurability is observed when $f_{\text{max}}$ is extended.

\section{Case Study: GW190814}\label{case0814}

The event GW190814, reported by the LIGO-Virgo-Kagra collaboration at a SNR of $\rho=25$ \cite{LIGOScientific:2020zkf}, consists of one black hole, weighting $22.2-24.3M_{\odot}$, and a compact object with $2.50-2.67M_{\odot}$. The mass of the latter intriguingly falls in the so-called ``lower mass gap'' ($2.5-5M_{\odot}$), and may be either the lightest black hole or the heaviest NS known to-date.
A possibility that the secondary is a mass-gap, fast-rotating NS has been raised in \cite{Zhang:2020zsc,Most:2020bba,Biswas:2020xna}, with the highest suggested spin being $\nus\sim 1170 \text{ Hz}$ \cite{Biswas:2020xna}. Although the spin parameter for this presumably, rapidly-rotating NS has not been well constrained, an estimation of the dimensionless spin $\chi$ via the relation (cf.~Eq.~(3) of \cite{Hannam:2013uu}),
\begin{align}
	\chi\approx0.4\bigg(\frac{\nus}{10^3\text{ Hz}}\bigg),
\end{align}
gives $\chi\approx0.47$ for the rate $\nus\sim 1170 \text{ Hz}$, which is about 65\% of maximum spin ($\chi\sim0.7$) attainable by an isolated NS \cite{Lo:2010bj}. This peculiar system may originate from a dynamical process, such as dynamical encounters in a star cluster \cite{Sedda:2020wzl,ArcaSedda:2021zmm}, hierarchical triple system \cite{Lu:2020gfh}, and tidal capture \cite{Mardling} of a natal NS kicked off from its born site by a BH.

Compared to the other tidal contributions, dephasing due to spin effects is secondary. Still, it seems that the estimation of spin parameter via its impact on the tidal dephasing may be promising. In this section, we discuss the tidally-induced phase shift for a fast-spinning NS, and estimate how can we probe both the $f$-mode frequency and the stellar spin rate from the waveform of GW190814 if the secondary turns out to be a fast-rotating NS.

\subsection{Estimation of Source Parameters}

\begin{figure}
	\centering
	\includegraphics[width=\columnwidth]{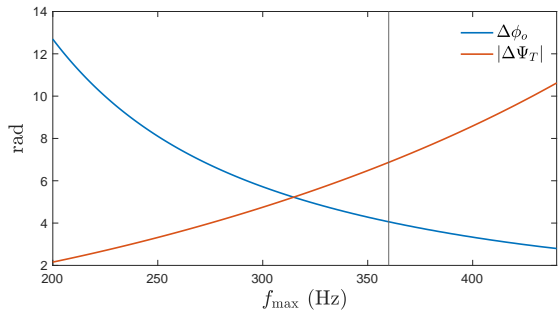}
	\caption{
	Same plot as the top panel of Fig.~\ref{fig:compare}, while the spin of the primary is $\nus=800\text{ Hz}$ here. The vertical line marks the merger frequency 360 Hz.
}
	\label{fig:compete}
\end{figure}

Adopting the definition of the onset of the merger as \cite{Yamamoto:2008js}, i.e., when $\pi(\Ms+\Mc)\gw=0.2$, the total mass of $\sim27M_{\odot}$ of GW190814 suggests that the merger occurred at $\gw\lesssim360\text{ Hz}$. In our simulation of the binary having two constituents with the masses and radii of those for GW190814, we find the separation between the two bodies is $\sim95$ km when $\gw=360\text{ Hz}$, which is larger than the sum of the two radii, viz.~$\sim75$ km.
We therefore set the cutoff at $f_{\text{max}}=360\text{ Hz}$ and the ``competition'' between $\Delta\Psi^T$ and $\Delta\phi_o$ is plotted in Fig.~\ref{fig:compete}. We see that the tidal dephasing in the waveform is exceeding the error of phase even with this low cutoff since the spin is rather high. Ignorance of the tidal effect in this case will therefore deteriorate the extraction of source parameters to an extent worse than ignoring the uncertainty in the reference phase $\phi_o$.

\begin{figure}
	\centering
	\includegraphics[width=\columnwidth]{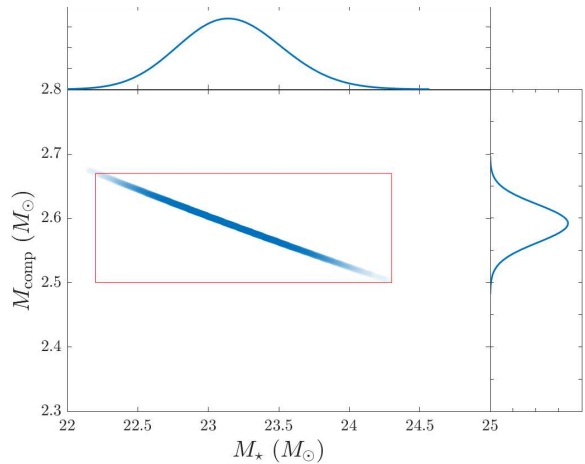}
	\caption{
	Posterior distribution of the primary and secondary source masses for the waveform model that assumes an aligned spin $\nus=800$~Hz, $f_{\text{max}}=360$~Hz, and EOS MPA1. The red rectangular plots the estimate reported in \cite{LIGOScientific:2020zkf}.
}
	\label{fig:bay}
\end{figure}

Although we show that the inclusion of tidal dephasing is necessary since the primary spins rapidly, the uncertainties for $\tilde{\Lambda}$ and $\omega_f$ are however large for this low $f_{\text{max}}$. We investigate the uncertainties of these two parameters as functions of $f_{\text{max}}$ in a neighbourhood of 360 Hz. In particular, we find the following relations
\begin{subequations}\label{eq:compete}
\begin{align}
    \frac{\Delta\omega_f}{\omega_f} = a_1 \left(\frac{f_{\text{max}}}{360 \text{ Hz}}\right)^{a_2} \%,
\end{align}
and
\begin{align}
    \frac{\Delta\tilde{\Lambda}}{\tilde{\Lambda}} = b_1 \left(\frac{f_{\text{max}}}{360 \text{ Hz}}\right)^{b_2} \%,
\end{align}
\end{subequations}
for the chosen EOS and $f_{\text{max}}\in[200,400]\text{ Hz}$, where the fitting parameters $a_i$ and $b_i$ are listed in Tab.~\ref{tab:compete}. The $b_2$ parameter is consistent with the trend showed in the Fig.~10 of \cite{Kawaguchi:2018gvj}.
For this specific system and $f_{\text{max}}=360$~Hz, we find additionally that the errors in ${\cal M}$ and $\eta$ are, respectively, $0.0066\%-0.0077\%$ and $2.787\%-2.813\%$ depending on the EOS. 
Assuming Gaussian priors for the chirp mass and the symmetric mass ratio centering at the peak of posterior distribution obtained in \cite{LIGOScientific:2020zkf}, i.e., ${\cal M}=6.09$ and $\eta=0.112$, we plot in Fig.~\ref{fig:bay} the probability distribution of our estimated of the masses of two objects, overlapped with their estimates (red rectangular). Although the estimates in \cite{LIGOScientific:2020zkf} are made without considering dynamical tides, the $\gtrsim 6$ rads dephasing [cf.~Fig.~\ref{fig:compete}], amounting to $0.33\%$ of the total cycles observed above $20$~Hz for GW190814, is not expected to affect much the inferences on ${\cal M}$ and $\eta$. However, ignoring $f$-mode excitation can deteriorate the estimate of $\tilde{\Lambda}$, which contributes only $\lesssim 1$ rad from 20 Hz to 360 Hz to the dephasing in Fig.~\ref{fig:compete}.

Although we assumed the knowledge of the spin in the above analysis, we note that we may estimate the spin by exploiting the universality of the tidal dephasing as a function of the dimensionless spin [Fig.~\ref{fig:uni_psi_spin}; Eq.~\eqref{eq:uni_psi_spin}], together with the mass measurement and the radius inferred by $\tilde{\Lambda}$ in the fashion of \cite{Raithel:2018ncd}. This will be considered elsewhere since the other spin effects, e.g., spin-orbit, spin-spin, and self-spin, should also be taken into account, which is beyond the scope of the present article.

We close the section by pointing out that if the secondary of GW190814 turns out to be an aligned NS rotating at $800$ Hz, the excitation of $f$-mode may generate a strong enough strain at $\sim 0.6$~s to crack the stellar surface in the fashion of \cite{Tsang:2011ad,Suvorov:2020tmk,Kuan:2021sin}.
Although there is no observation of electromagnetic counterparts for GW190814, this does not necessary falsify the possibility that the secondary is a NS. The absence of electromagnetic signal might be because the emission is off-beam, absorbed by the black hole remnant, or the surface strength of magnetic field might be weaker than the critical value. For example, the non-detection of a precursor for GW170817 gives an upper bound on the magnetic fields of both NS progenitors of $\sim 10^{13.5}$ G assuming the resonance of interface mode is responsible for crustal failure \cite{neil22}. 

\begin{table}
	\centering
	\caption{Parameters relevant to Eq.~\eqref{eq:compete} for the chosen EOS. The inertial-frame, spin-modified $f$-mode frequency is listed in the second column, and the fitting parameters defined in Eq.~\eqref{eq:compete} are presented from the third to the final column. Here we assume $\nus=800\text{ Hz}$.}
	\begin{ruledtabular}
	\begin{tabular}{cccccc}
		EOS & $\omega_f/2\pi$ (Hz) & $a_1$ & $a_2$ & $b_1$ & $b_2$ \\
		\hline 
		KDE0V & 817.57 & 509.55 & -2.46 & 9.05 & -2.34 \\
		APR4 & 789.65 & 436.70 & -2.49 & 7.86 & -2.36 \\
		SLy & 748.47 & 321.41 & -2.53 & 5.91 & -2.38 \\
		ENG & 697.92 & 241.25 & -2.60 & 4.59 & -2.42 \\
		MPA1 & 637.01 & 151.46 & -2.70 & 3.04 & -2.47
	\end{tabular}
	\end{ruledtabular}
	\label{tab:compete}
\end{table}

\section{Discussion} \label{discussion}

The phase of gravitational waveform is sensitive to several stellar parameters, which elevates it into an invaluable position in the GW physics era. Among other factors, the tidal contribution to the GW phasing encodes the details of internal motions of NSs, consisting of the equilibrium tide, described by the tidal deformability (Sec.~\ref{analtidphi}), and the dynamical, $f$-mode oscillation, captured by the mode frequency $\omega_f$ and its coupling strength to the external tidal field $Q_f$ [Eq.~\eqref{eq:tidhim}].
The determination of $\Lambda$ via the phase shift due to equilibrium tides can set constraints on the EOS as demonstrated by the analysis of GW 170817 \cite{LIGOScientific:2018cki}. However, we need to take also into account the dynamical tides since more accurate observations will be available in the near future. 
This entails a good handle on the QNM effects on the waveforms especially in the high frequency window, where the influence of tidal effects in the GW signal is encoded (top panel of Fig.~\ref{fig:spin}). 
Ignoring the dynamical tide contribution in the phase shift will therefore deteriorate the accuracy in constraining the EOS. Furthermore, for binaries involving rapidly-rotating NSs the effect will be more pronounced since the $f$-mode frequency will be lowered, leading into larger tidal dephasing (Sec.~\ref{stat.err}).

As we mention earlier, both spin and tidal effects will influence the GW phasing though the spin contribution is smaller \cite{Dietrich:2016lyp}. Therefore, it is crucial to estimate the phase shift caused by each one of them for the precise estimation of the tidal parameters (e.g., \cite{Agathos:2015uaa}). 
If we can determine independently the stellar spin through, e.g., the range of dynamical ejecta \cite{Most:2019pac,Ruiz:2019ezy}, a shift in the main pulsating mode in hypermassive NS remnant \cite{Dietrich:2016fpt,East:2019lbk}, or a system showing double precursors \cite{double}, we can construct a point-particle waveform for that spin. By subtracting the point-particle part of the waveform from the data, we can get the tidal dephasing. 
In addition, the tidal effects are encoded in the high frequency part of GW data stream, while the spin affects mainly the low frequency part \cite{Harry:2018hke,Dietrich:2020eud}. 
Thus we may acquire the individual spins in the early part of the waveform by measuring spin-orbit and spin-spin contributions. Although the latter spin-spin contribution is degenerate with the self-spin effect as discussed in Introduction, the I-Love-Q relation can help in breaking the degeneracy since the spin-induced quadrupole moment can be estimated from the adiabatic tidal parameter \cite{Yagi:2013bca,Yagi:2013awa}.
On the other hand, GW luminosity during inspiral is more sensitive to the tidal effects than the spin-orbit terms \cite{Zappa:2017xba}. In particular, the tidal contribution of the primary to the luminosity reads \cite{Hinderer:2009ca}
\begin{align}
	\mathcal{L}_{\text{GW}}^{T} = \frac{192\eta^2x^{10} (\Ms+3\Mc)\Ms^4\Ls}{5(\Ms+\Mc)^5},
\end{align}
while the spin-orbit contribution is negligible. Therefore, by correlating the tidal imprints in the phasing and the luminosity, we may provide more accurate estimation of the tidal parameters. Nonetheless, we note that the above equation applies to non-spinning NSs.

In reality, the tidal effect on the members of NS binaries can be 3-fold: \emph{gravito-electric}, \emph{gravito-magnetic} tides, and the \emph{change in the waveform} shape induced from the gravito-electric tidal field \cite{Damour:2009vw,Poisson:2020eki,Poisson:2020mdi}. 
Our focus in the present article was on the gravito-electric tides, while we should keep in mind that the excitations by the gravito-magnetic tidal field \cite{Ma:2020oni,Gupta:2020lnv} may become comparable to the gravito-electric excitations under certain conditions \cite{Poisson:2020ify}. Although we do not consider spin contribution in the GW dephasing since it is minor compared to the dephasing by equilibrium tides  (e.g., \cite{Dietrich:2016fpt}), we are aware that the uncertainty in the spin contribution will affect, to certain extent, the estimation of the tidal parameters; detailed analysis of tidal dephasing, where the spin-induced dephasing is included, would be useful.
In addition, during the preparation of this article, a new study pointing out the importance of tidal-spin interaction in the waveform modelling for fast spinning NS with $\chi\gtrsim0.1$ was published \cite{Castro:2022mpw}.

\section*{Acknowledgement}
KK gratefully acknowledges financial support by DFG research Grant No. 413873357. 
HJK is indebted to the support from Sandwich grant (JYP) No. 109-2927-I- 007-503 by DAAD and MOST during early stages of this work. The authors thank the anonymous referees for their useful feedback, which improves the quality of this work.

\section*{Data availability statement}
Observational data used in this paper are quoted from the cited works. Data generated from computations are reported in the body of the paper. Additional data will be made available upon reasonable request.


\bibliographystyle{apsrev4-2}
\bibliography{references}

\end{document}